\documentclass[superscriptaddress,groupedaddress,nofootnoteinbib,12pt]{article}
\usepackage{color}
\usepackage{graphicx}
\usepackage{dcolumn}
\usepackage{bm}
\usepackage{amssymb}
\usepackage{amsmath}
\usepackage{sectsty}
\usepackage{colortbl}
\usepackage{latexsym}
\usepackage{float}
\usepackage{ifthen}
\usepackage{caption,subfig}
\usepackage{enumerate}
\usepackage{url}
\usepackage{caption,subfig}
\usepackage{feynmf}	
\usepackage{jcappub}

\usepackage{enumitem}
\newlist{Properties}{enumerate}{2}
\setlist[Properties]{label={\color{blue}Property \arabic*.},leftmargin=*,ref={\color{blue}Property\  \arabic*}}

\def\be{\begin{equation}}
\def\ee{\end{equation}}
\def\ba{\begin{eqnarray}}
\def\ea{\end{eqnarray}}
\def\beq{\begin{eqnarray}}
\def\eeq{\end{eqnarray}}

\def\mpl{M_{\rm Pl}}

\def\E{\mathcal{E}}

\def\d{\mathrm{d}}
\def\p{{\cal P}}

\def\K{{\cal K}}

\def\L*{{\cal L}_*}
\def\L{\mathcal{L}}
\def\({\left(}
\def\){\right)}
\def\ie{{\it i.e. }}

\def\nn{\nonumber}
\def\p{\partial}
\def\mn{_{\mu \nu}}
\def\stu{St\"uckelberg }
\def\p{\partial}
\def\mupn{^\mu_{\ \nu}}
\def\<{\langle}
\def\>{\rangle}
\def\Ein{\hat{\mathcal{E}}}

\def\pa {\partial}

\def\cs2{c_{s}^{2}}

 \def\ep{\varepsilon}

 \def\Om{\Omega}
 
 \def\p{\partial}

\def\Lk{\mathcal{L}^{\rm (der)}}
\def\Lm{\mathcal{L}^{\rm (m)}}
\def\Lkb{\bar{\mathcal{L}}^{\rm (der)}}
\def\Lmb{\bar{\mathcal{L}}^{\rm (m)}}

\def\mpl{M_{\rm Pl}}
 \def\be   {\begin{equation}}   \def\ee   {\end{equation}}

 \def\ba  {\begin{eqnarray}}   \def\ea  {\end{eqnarray}}

\begin{document}

\title{New Kinetic Interactions for Massive Gravity?}

\author{Claudia de Rham, Andrew Matas,}
\author{and Andrew J. Tolley}
\affiliation{CERCA/Department of Physics, Case Western Reserve University, 10900 Euclid Ave, Cleveland, OH 44106, USA}
%\date{\today}

%%%%%%%%%%%%%%%%%%%%%%%%%%%%%%%%%%%%%%%%%%%%%%%%%%%%%%%%%%%%%%%%%%%%%
%%%% Abstract

%Massive Gravity in four dimensions has been shown to be free of the Boulware-Deser (BD)  ghost
%both in the ADM and the \stu languages for a  specific choice of mass terms.  We show here how this is consistent with the
%helicity language.
\abstract{We show that there can be no new Lorentz invariant kinetic interactions free from the Boulware-Deser ghost in four dimensions in the metric formulation of gravity, beyond the standard Einstein-Hilbert, up to total derivatives.
We use dimensional deconstruction as a way to motivate a non-linear ansatz for potential new ghost free kinetic interactions for massive gravity, bi-gravity and multi-gravity in four and higher dimensions. These interactions descend from Lovelock terms, and so naively one might expect the interactions to be ghost free. However we show that these new interactions inevitably lead to more than five propagating degrees of freedom. We then perform a general perturbative analysis in four dimensions, and show that the only term with two derivatives that does not introduce a ghost is the Einstein-Hilbert term. This result extends to all orders in perturbations.}

\maketitle

%%%%%%%%%%%%%%%%%%%%%%%%%%%%%%%%%%%%%%%%%%%%%%%%%%%%%%%%%%%%%%%%%%%%%
%%%% Introduction

\section{Introduction}
The kinetic terms of theories with spin greater than zero are generically tightly constrained by the requirements of Lorentz invariance and unitarity of the quantum theory (or stability of the classical theory). For example, it is well known for massive electromagnetism that even after the gauge invariance is broken by the mass term, the kinetic term must remain gauge invariant to prevent the existence of ghosts \cite{Fierz:1939ix}. \\

However, it was shown in \cite{Hinterbichler:2013eza} (see also \cite{Folkerts:2011ev,Folkerts:2013mra} and \cite{Kimura:2013ika}) that at the linear level around a Minkowski background there are non-gauge invariant ghost-free kinetic terms for a spin-2 field, and it was conjectured that there may be a non-linear diffeomorphism-violating completion that avoids the Boulware-Deser (BD) ghost \cite{Boulware:1973my} when coupled to matter.\footnote{Note that, like dRGT massive gravity, the goal is to construct a gravitational theory that propagates five degrees of freedom around every background, and so is free of the sixth Boulware-Deser mode. However, there may be pathological backgrounds where one of the five propagating degrees of freedom becomes ghost-like (for examples, see \cite{Fasiello:2012rw,Khosravi:2013axa,DeFelice:2013awa}). We emphasize that we are not addressing existence of pathological backgrounds in this work, we are only concerned with the Boulware-Deser ghost.} The existence or nonexistence of non-standard BD-ghost-free kinetic terms has major implications for massive gravity, bi-gravity, and multi-gravity. For example, it was shown in \cite{deRham:2013wv} that new kinetic terms would be required for the existence of partially massless gravity (for more on partially massless gravity see \cite{deRham:2012kf,Deser:2012qg,Deser:2013xb,Deser:2013bs,Deser:2013uy}). Additionally, new kinetic interactions, if they existed, could change the form of the Hamiltonian constraint in the minisuperspace, possibly allowing for the existence of exact FRW solutions in massive gravity \cite{D'Amico:2011jj}. Finally, if massive gravity could allow for new kinetic interactions, it might lead to modifications of gravity not only in the infrared. \\

Furthermore it was shown in \cite{deRham:2013awa} that the BD-ghost-free \cite{deRham:2010ik,deRham:2010kj,deRham:2011rn,Mirbabayi:2011aa,Hassan:2011hr,Hassan:2011ea} dRGT mass terms can be derived from a five dimensional Einstein-Hilbert term using the method of Dimensional Deconstruction. There it was found that it was crucial to discretize the vielbein, and not the metric. See \cite{ArkaniHamed:2002sp,ArkaniHamed:2003vb,Schwartz:2003vj,Deffayet:2005yn,Deffayet:2003zk,Deffayet:2004ws} for other work on applying Dimensional Deconstruction to gravity on a flat compact dimension. But there is one other ghost free kinetic term in five dimensions: the Gauss-Bonnet term! Thus it is extremely plausible that if there are new kinetic interactions in massive gravity, they can be derived from Dimensional Deconstruction. Conversely, if Dimensional Deconstruction applied to Gauss-Bonnet causes the constraints present in the continuum theory to be lost, it seems unlikely that there could be another combination of non-standard kinetic terms that propagates five degrees of freedom. \\

We will show that Dimensional Deconstruction applied to Gauss-Bonnet produces interactions that propagate more than five degrees of freedom when discretized using the prescription in \cite{deRham:2013awa}. However, the derivative structure of Gauss-Bonnet is more intricate than that of Einstein-Hilbert, and so it is more sensitive to the process of discretization. In a generic theory involving interactions of quadratic or higher order in the Riemann curvature, such as $R_{\mu\rho\rho\sigma}^2$, the theory will contain an Ostragradski ghost since the action contains higher derivative interactions $R^2 \sim (\partial^2 h)^2$. The Gauss-Bonnet term is special precisely because the dangerous higher derivative interactions arise as a total derivative. By discretizing the $y$ derivative, we break the total derivative structure, and make the higher derivative pieces physical. The failure of the deconstruction procedure is already evidence that new kinetic interactions do not exist.\\

We further perform a perturbative analysis to identify potentially new BD-ghost-free kinetic terms. We find that any new kinetic interaction must be identical to the Einstein-Hilbert term up to quartic order. The absence of new terms beyond quartic order is easily established since any such term should at leading nontrivial order in perturbations be BD-ghost-free when introducing \stu fields for linearized diffeormorphisms (diffs). However, it has already been established that there are no new kinetic terms that satisfy this criterion beyond cubic order \cite{Hinterbichler:2013eza}, and consequently there can be no nonlinear completion of such terms. As a result, \textbf{we prove to all orders in perturbations that there can be no other Lorentz covariant kinetic term for massive gravity than the standard Einstein-Hilbert one}, up to total derivatives. The results presented in this paper agree with the special cases considered in \cite{Kimura:2013ika}. \\

The rest of the paper is organized as follows: in section~\ref{sec:linear-diff} we shall review the reasoning behind conjecturing that there could be interactions with linearized diffs. In section~\ref{sec:deconstruction} we shall apply Dimensional Deconstruction to the Gauss Bonnet term in five dimensions as a physically motivated ansatz for a non-linear completion for these terms. We shall review how Dimensional Deconstruction can be applied to the five-dimensional Einstein-Hilbert term to produce BD-ghost-free massive gravity, bi-gravity, and multi-gravity. However, the resulting interactions can easily be shown to have ghosts already in the minisuperspace approximation. Thus in section~\ref{sec:no-go} we perform a brute force perturbative analysis, and show that up to quartic order
the only term that propagates no more than five degrees of freedom is the Einstein-Hilbert term. This rules out any new kinetic interaction {\it to all orders} in the metric language. The only allowed terms with more than two derivatives must be total derivatives.

%
%
%%%% Section 2: Review of Potential and Kinetic Interactions at Leading Order
%
%
\section{Review of Massive and Kinetic Interactions}
\label{sec:linear-diff}
General Relativity is the interacting theory of a massless spin 2 particle. This means the theory must propagate 2 helicity-2 degrees of freedom around an arbitrary background in 4 space-time dimensions. This requirement is strong enough to completely fix the action for pure gravity (see for example \cite{Gupta:1954zz,Weinberg:1965rz,Deser:1969wk,Feynman:1996kb,Boulware:1974sr}). The only allowed term with no derivatives is the cosmological constant. The only allowed term with two derivatives on the metric is the Einstein-Hilbert term.

\subsection{Terms allowed by the linearized \stu decomposition}

Similarly, the defining feature of massive gravity is that it propagates the five degrees of freedom of a massive spin 2 particle. This constraint is again very powerful, and only a handful of interactions are allowed. This was shown already at the linearized level by Fierz and Pauli, \cite{Fierz:1939ix} and was extended non-linearly in \cite{deRham:2010ik,deRham:2010kj}.

\subsubsection{Fierz-Pauli Theory}
\label{sec:FPtheory}

Focusing for a moment at the linear level, it has was shown by Fierz and Pauli that the theory for a {\it non-interacting} massive spin-2 field $H$ has to be
\ba
\label{FP}
S_{\rm FP}&=&\int \d^4 x\(-\frac{\mpl^2}4  H^{\mu\nu}\Ein^{\alpha\beta}\mn  H_{\alpha\beta}+\frac{m^2\mpl^2}{4}\Lm_{2}\)\,,
\ea
where $\Ein$ is the Lichnerowicz operator which corresponds to the linearized Einstein-Hilbert kinetic term and $\Lm_{2}$ is the Fierz-Pauli mass term,
\ba
\label{Lichnerowicz operator}
\Ein^{\alpha\beta}\mn  H_{\alpha\beta}&=&-\frac 12 \(\Box  H\mn-2\p_{(\mu}\p_\alpha   H^\alpha_{\nu)}+\p_\mu\p_\nu  H-\eta\mn (\Box  H-\p_\alpha\p_\beta  H^{\alpha\beta})\)\,,\\
\label{FP mass term}
\Lm_{2}&=& \frac{1}{2}\(H^2-H\mn^2\)\,.
\ea
At this level $H$ is a non-interacting  massive spin-2 field living on flat space-time, and so \eqref{FP} is not a theory of gravity. All indices are raised and lowered with respect to the flat Minkowski metric. The kinetic term $H \Ein   H $ is invariant under {\it linear} diffeomorphisms, but the mass term breaks that symmetry. We can nevertheless restore it by use of the  {\it linear} \stu decomposition
\ba
\label{eq:linear-stuck}
H_{\mu\nu} = \frac{h_{\mu\nu}}{\mpl} + \frac{1}{m \mpl}\left(\pa_{\mu} A_\nu + \pa_\nu A_\mu \right) + \frac{2}{m^2 \mpl} \pa_\mu \pa_\nu \pi\,,
\ea
so that the theory is invariant under linear diffs and under an additional $U(1)$ symmetry,
\ba
\delta_\xi h\mn =\p_\mu \xi_\nu+ \p_\nu \xi\,, \hspace{30pt} \delta_\xi A_\mu =- m \xi_\nu \\
\delta_\Lambda A_\mu= \p_\mu \Lambda\,, \hspace{30pt} \delta_\Lambda \pi = - m \Lambda\,.
\ea
The \stu decomposition allows us to identify the degrees of freedom: $h\mn$ represents the helicity-2 mode (2 dofs), $A_\mu$ the helicity-1 mode (2 dofs) and $\pi$ the helicity-0 mode (1 dof).
The linearized Einstein-Hilbert kinetic term is of course insensitive to that decomposition $H \Ein H = h \Ein h$, and the helicity-0 and -1 modes  $A$ and $\pi$ only appear in the mass term. The combination that enters in the Fierz-Pauli mass term \eqref{FP mass term} is special in that it does not generate more than two derivatives on any of the fields when performing the linearized \stu decomposition (this will be refered as \ref{Prop lin Stuc}).
 Any other combination would have led to a term going as $(\Box \pi)^2$ which by Ostrogradsky's theorem \cite{Ostrogradsky} would have signaled the presence of a ghost, (this is typically known as the Boulware-Deser (BD) ghost, \cite{Boulware:1973my}  see also \cite{ArkaniHamed:2002sp,Creminelli:2005qk,Deffayet:2005ys}).

\subsubsection{New mass terms starting at higher order}
\label{sec:NewMassTerms}
In principle one could be also include new terms which respect the same following property:
\begin{Properties}
\item  \label{Prop lin Stuc}
\emph{At leading order} in that new term, it must be free of any higher derivatives when performing the  \emph{linearized} \stu decomposition \eqref{eq:linear-stuck}.
\end{Properties}
The Fierz-Pauli mass term is actually not the only potential term that satisfies the property,
in four dimensions,  there are two additional contributions which are respectively cubic and quartic order in $H$, (we include the Fierz-Pauli mass term for comparison and completeness)
\ba
\label{lm2lin}
\Lm_2 &=& \frac{1}{2^2}\ep^{\mu \nu \alpha \beta} \ep_{\mu' \nu' \alpha \beta} H^{\mu'}_{\, \mu} H^{\nu'}_{\, \nu} \\
\label{lm3lin}
\Lm_3 &=& \frac{1}{2^3}\ep^{\mu \nu \rho \alpha} \ep_{\mu' \nu' \rho' \alpha} H^{\mu'}_{\, \mu} H^{\nu'}_{\, \nu} H^{\rho}_{\, \rho'} \\
\label{lm4lin}
\Lm_4 &=& \frac{1}{2^4}\ep^{\mu \nu \rho \sigma} \ep_{\mu' \nu' \rho' \sigma'} H^{\mu'}_{\, \mu} H^{\nu'}_{\, \nu} H^{\rho}_{\, \rho'}H^\sigma_{\, \sigma'}\,,
\ea
where $\ep$ is the Levi-Cevita symbol ($\ep=\{0,1,-1\}$ and does not depend $H$). One can easily see that when performing the linear \stu decomposition \eqref{eq:linear-stuck}, none of these terms lead to any higher derivatives in any of the fields. This statement is {\it exact} and does not rely on any decoupling limit analysis or other approximation. It has been shown that these three potentials for $H$ were the only one satisfying this property \cite{deRham:2010ik,deRham:2010kj} and there are no other potential term that  does not lead to an Ostrogradsky instability for one of the \stu  fields introduced by performing the linear \stu decomposition \eqref{eq:linear-stuck}.

One important subtlety is that only $\Lm_2$ can generate a kinetic term for the \stu, and so to avoid having infinitely strongly coupled degrees of freedom, we cannot include $\Lm_3$ or $\Lm_4$ without also including $\Lm_2$.

\subsubsection{New kinetic terms starting at higher order}
\label{sec:NewKineticTerms}
Just like we were able to find `new' mass terms at higher order for the Fierz-Pauli theory, one can try the same endeavour for the kinetic terms. Could there be other allowed derivative interactions in the theory of a pure massive spin-2 field (\ie not a theory of gravity)?

As with the mass terms, a necessary condition for a potentially new kinetic interaction is given in \ref{Prop lin Stuc}.
Such a term was identified recently in \cite{Hinterbichler:2013eza}\footnote{This term was actually identified previously in \cite{Folkerts:2011ev} but using a helicity decomposition argument which fails at higher order, see \cite{deRham:2011qq}.} at cubic order in four dimensions (we include the Einstein-Hilbert term $\Lk_{2}$ for comparison and completeness),
\ba
\Lk_{2}&=&H^{\mu\nu}\Ein^{\alpha\beta}\mn  H_{\alpha\beta} = \frac 12
\ep^{\mu \nu \rho \sigma} \ep^{\mu' \nu' \rho' \sigma'} \eta_{\sigma\sigma'}\p_\rho H_{\mu\mu'} \p_{\rho'}H_{\nu \nu'}  \\
\Lk_{3}&=& \ep^{\mu \nu \rho \sigma} \ep^{\mu' \nu' \rho' \sigma'} H_{\sigma\sigma'}\p_\rho H_{\mu\mu'} \p_{\rho'}H_{\nu \nu'}\,.
\ea
This procedure can be generalized to an arbitrary number of dimensions, (with more terms in higher dimensions). In four space-time dimensions it was shown that $\Lk_{2,3}$ were the unique terms that satisfy the required \ref{Prop lin Stuc} \cite{Hinterbichler:2013eza}.

Here again notice that we cannot include $\Lk_3$ without also including $\Lk_2$. Furthermore as mentioned earlier the \stu fields do not enter the Einstein-Hilbert term $\Lk_2$ and their kinetic term arise from $\Lm_2$. Since the \stu do enter $\Lk_3$, one cannot consider $\Lk_3$ without also including $\Lm_2$.

Such an analysis can only tell us about the leading order behavior in an expansion of a non-linear interaction. We now seek to go beyond the leading order approximation and see if there exists a non-linear completion to these derivative interactions.

 %that in $D$ space-time dimensions the following class of derivative interactions beyond the standard Einstein-Hilbert were identified that satisfy this condition. A term with $d$ derivatives and $n_h$ powers of $h$  is given schematically by
%
%\be
%\mathcal{L}_{d,n_H} \sim \ep \ep (\pa H \pa H)^{d/2} H^{n_H-d} \eta^{D-n_H-d/2}
%\ee
%Clearly $d$ must be even in order for the Lagrangian to be a scalar. Additionally, we must have $ n_H + d/2 \leq D$ and $d \leq n_H$. In $D=4$ the only possibilities with nonzero $d$ are given by $d=2$ and $n_H = 2,3$. Explicitly these terms are
%\ba
%\label{eq:L_dn}
%\mathcal{L}_{2,2} &=& \ep^{\mu\nu\rho\sigma}\ep^{\mu' \nu' \rho' \sigma'}\pa_\mu H_{\nu \nu'}\pa_{\mu'} H_{\rho \rho'} \eta_{\sigma \sigma'} \nn \\
%\mathcal{L}_{2,3} &=& \ep^{\mu\nu\rho\sigma}\ep^{\mu' \nu' \rho' \sigma'}\pa_\mu H_{\nu \nu'} \pa_{\mu'} H_{\rho \rho'} H_{\sigma \sigma'}.
%\ea
%The term $\mathcal{L}_{2,2}$ is identical to the familiar Einstein-Hilbert action at quadratic order. However, it is possible that there exists a non-linear kinetic interaction that is degenerate with the Einstein-Hilbert term at the quadratic level, but differs at higher order. The existence of such a term cannot be deduced by considering only the leading order behavior. \\

\subsection{Non-linear completions}

So far the Fierz-Pauli theory (be it implemented with the `new' terms $\Lm_{3,4}$ and $\Lk_{3}$ or not) is the theory for a massive spin-2 field which does not interact with anything else. If we would like this theory to be relevant for gravity, the spin-2 field needs to interact with matter and $H\mn$ is part of a dynamical metric, for instance $g\mn=\eta\mn+H\mn$. Linearized diffeomorphism should then be traded for general coordinate invariance \ie diffeomorphism invariance.

Rather than introducing the linearized \stu decomposition which would restore linearized diffeomorphism invariance, we should thus consider the full {non-linear} \stu decomposition \cite{Siegel:1993sk,ArkaniHamed:2002sp}
\ba
\label{eq:non-linear-stuck}
H_{\mu\nu}&=&\frac{h_{\mu\nu}}{\mpl}+ \frac{1}{m \mpl}\partial_{(\mu}A_{\nu)} + \frac{2}{m^2 \mpl} \pa_\mu \pa_\nu \pi \nn \\
&&-\eta^{\rho \sigma} \left( \frac{\partial_{\mu} A_{\rho}}{m \mpl}+  \frac{\pa_\mu \pa_\rho \pi}{m^2 \mpl}\right)\left( \frac{\partial_{\sigma} A_{\nu}}{m \mpl}+  \frac{\pa_\sigma \pa_\nu \pi}{m^2 \mpl}\right).
\ea
where $(a,b)=ab+ba$. We could consider another metric but restrict ourselves to Minkowski for simplicity.

The physical relevance of this non-linear \stu decomposition is that upon taking the decoupling limit $m\rightarrow 0, \mpl \rightarrow \infty$ with $\Lambda_3 \equiv (m^2 \mpl)^{1/3}$ fixed, the modes decouple from one another and live on the flat reference metric. For this reason we may once again identify $h_{\mu\nu}$ as the helicity-2 mode, $A_\mu$ as the helicity-1 mode, and $\pi$ as the helicity-0 mode (see \cite{deRham:2011qq} and \cite{deRham:2011rn} for more detailed discussions). So in order for an interaction to be BD-ghost-free it must also satisfy the following property:
\begin{Properties}
\setcounter{Propertiesi}{1}
\item If at leading order, a term satisfies \ref{Prop lin Stuc} then \textbf{in the decoupling limit} its {\it non-linear extension} must be free of any higher derivatives on any of the fields when performing the {\it non-linear} \stu decomposition \eqref{eq:non-linear-stuck}.
    \label{Prop DL}
\end{Properties}
This is a necessary condition for the absence of the BD ghost but not always a sufficient one. From the very nature of the non-linear \stu decomposition \eqref{eq:non-linear-stuck}, it is clear that the leading order terms by themselves cannot respect \ref{Prop DL} and should thus be extended or rather completed fully non-linearly. This should come at no surprise since we are used to dealing with fully non-linear interactions in General Relativity.

\subsubsection{Non-linear completion for the mass terms}
\label{sec:NonLinearMassTerms}
This \ref{Prop DL} is what {\it uniquely} fixes the fully non-linear completion of the mass terms \eqref{lm2lin}, \eqref{lm3lin} and \eqref{lm4lin}, (see Ref.~\cite{deRham:2010ik} and Ref.~\cite{deRham:2014zqa} for a review)
\ba
\label{eq:mass-terms}
\begin{array}{ccl}
\text{Leading order}\hspace{20pt} & &\hspace{20pt} \text{Fully non-linear realization}\\[5pt]
\hline \\[1pt]
\Lm_2 & \longrightarrow\hspace{20pt}&\hspace{20pt} \Lmb_2 =\sqrt{-g}\ \E^{\mu \nu \alpha \beta} \E_{\mu' \nu' \alpha \beta} \K^{\mu'}_{\, \mu} \K^{\nu'}_{\, \nu}  \\
\Lm_3 & \longrightarrow\hspace{20pt}&\hspace{20pt} \Lmb_3 = \sqrt{-g}\ \E^{\mu \nu \rho \alpha} \E_{\mu' \nu' \rho' \alpha} \K^{\mu'}_{\, \mu} \K^{\nu'}_{\, \nu} \K^{\rho}_{\, \rho'} \\
\Lm_4 & \longrightarrow\hspace{20pt}&\hspace{20pt} \Lmb_4 = \sqrt{-g}\ \E^{\mu \nu \rho \sigma} \E_{\mu' \nu' \rho' \sigma'} \K^{\mu'}_{\, \mu} \K^{\nu'}_{\, \nu} \K^{\rho}_{\, \rho'}\K^\sigma_{\, \sigma'}\,,
\end{array}
\ea

These three families of terms are fully non-linear for instance $\Lmb= H^2+ H^3+\cdots $ and contain an infinity of terms but we keep the subscripts  to indicate at which order that term starts. Here again we cannot consider $\Lmb_{3,4}$ without also including $\Lmb_2$.

In the previous expressions, $\E$ is the Levi-Cevita tensor and the tensor $\K$ is given by
\be
\K^\mu_{\, \nu} \equiv \delta^\mu_\nu - \left(\sqrt{g^{-1} f}\right)^\mu_{\ \nu}\,,
\ee
where as mentioned previously, we take the reference metric $f_{\mu\nu}=\eta_{\mu\nu}$ for simplicity and $H\mn$ is part of the dynamical metric
\ba
g_{\mu\nu} = \eta_{\mu\nu} + H_{\mu\nu}\,.
\ea
As mentioned previously, \ref{Prop DL} is a necessary condition for the absence of a BD ghost but not always a sufficient one. A more general requirement is given by the following condition
\begin{Properties}
\setcounter{Propertiesi}{2}
\item If at leading order, a term satisfies \ref{Prop lin Stuc} then its {\it non-linear extension} must be such that when performing an ADM decomposition, the Lagrangian must be put in a first order form which involves no time derivatives neither on the shift $N^i$ nor on the lapse $N^0$. Furthermore the determinant of the Hessian $L\mn=\delta^2 \mathcal{H}/\delta N^\mu \delta N^\nu$ must vanish, (where $\mathcal{H}$ is the Hamiltonian).
    \label{Prop ADM}
\end{Properties}
\ref{Prop ADM} is in fact a necessary and sufficient condition to guarantee that 5 or fewer degrees of freedom propagate (although it does not guarantee that all of them are healthy). It implies \ref{Prop DL}, however it is also computationally more difficult to check and is also more physically obscure.

This property ensures that the shift and the lapse do not enter the phase space (\ie only the spatial part of the metric and its conjugate enter, leading to 12 phase space variables) and that a combination of the shift and the lapse propagates a constraint reducing the phase space to 10 variables corresponding to five physical degrees of freedom. This argument was first formulated in \cite{deRham:2010ik} and carried out for the mass terms \eqref{eq:mass-terms} in \cite{deRham:2010kj} and \cite{Hassan:2011hr,Hassan:2011ea}. The proof for the absence of a BD ghost for the three possible mass terms \eqref{eq:mass-terms} was also carried through in the \stu language \cite{deRham:2011rn,Hassan:2012qv} and the connection with \ref{Prop DL} was established in  \cite{Mirbabayi:2011aa}. In the case of these mass terms, \ref{Prop DL} is indeed sufficient and ensures \ref{Prop ADM} but as we shall see below this is not the case for the kinetic terms.

\subsubsection{Non-linear completion for the kinetic terms}

The same non-linear completion can be undertaken for the kinetic terms.
%It is well known that $\Lk_2$ extends fully non-linearly to the Einstein-Hilbert term, or at least this is one possible realization which satisfies both\footnote{It actually satisfied \ref{Prop DL} trivially since the \stu fields do not enter the full Einstein-Hilbert term, since it is covariant. } \ref{Prop DL} and \ref{Prop ADM},
The most obvious fully non-linear extension to $\mathcal{L}^{\text{(der)}}_2$ is that satisfies \ref{Prop DL} and \ref{Prop ADM} is the Einstein-Hilbert term.\footnote{The Einstein-Hilbert term actually satisfies \ref{Prop DL} trivially since the \stu fields do not enter the full Einstein-Hilbert term, since it is covariant. } However, we can raise the question of whether or not this completion is unique.  \\
\noindent Furthermore in \cite{Hinterbichler:2013eza} it was conjectured that $\Lk_3$ is in fact the leading order piece of a fully non-linear term $\Lkb_3$  that propagates five degrees of freedom and should thus satisfy \ref{Prop DL} and \ref{Prop ADM}. The situation can be summarized as

%\noindent Furthermore in \cite{Hinterbichler:2013eza} it was conjectured that $\Lk_3$ is in fact the leading order piece of a fully non-linear term $\Lkb_3$  that propagates five degrees of freedom and should thus satisfy \ref{Prop DL} and \ref{Prop ADM},
%\ba \label{kin extensions}
%\begin{array}{ccc}
%\text{Leading order}\hspace{20pt} & &\hspace{30pt} \text{Fully non-linear completion}\\[5pt]
%\Lk_2 & \hspace{30pt}\longrightarrow\hspace{30pt}&\hspace{30pt} \Lkb_2 =\sqrt{-g}\ R[g]  \\
%&\hspace{30pt}\longrightarrow\hspace{30pt}&\hspace{30pt} \text{Other non-linear completions ?}  \\[3pt]
%\Lk_3 &\hspace{30pt} \longrightarrow\hspace{30pt}&\hspace{30pt} \Lkb_3= \ ?
%\end{array}
%\ea

\ba \label{kin extensions}
\begin{array}{ccc}
\text{Leading order}\hspace{20pt} & &\hspace{30pt} \text{Fully non-linear completion}\\[3pt]
\hline \\
\Lk_2 & \hspace{30pt}\longrightarrow\hspace{30pt}&\hspace{30pt}
\begin{cases}
 \Lkb_2 =\sqrt{-g}\ R[g]  \\
  \text{Other non-linear completions ?}
  \end{cases} \\[5pt]
\Lk_3 &\hspace{30pt} \longrightarrow\hspace{30pt}&\hspace{30pt} \Lkb_3= \ ?
\end{array}
\ea

In this paper we shall see that the only completion of the linearized Einstein-Hilbert term is the Einstein-Hilbert term\footnote{This statement is more profound than it may seem: Requiring that no more than five degrees of freedom propagate in the kinetic term we deduce that only the Einstein-Hilbert term is acceptable, which is covariant and by itself only propagates two degrees of freedom.} and that there is no possible completion to the kinetic term $\Lk_3$.
In order to gain physical insight into the origin of such a term $\Lkb_3$ or other completions to $\Lk_2$, we now turn to higher dimensions and apply dimensional deconstruction to the Lovelock terms in five dimensions.

%
%
%%%% Section 3: Kinetic Interactions from Deconstruction
%
%
\section{Kinetic Interactions from Deconstruction}
\label{sec:deconstruction}
\subsection{Deconstruction and Massive Gravity}
\label{sec:deconstructing-eh}

First  we review how dimensional deconstruction can be used to generate the BD-ghost-free dRGT mass terms \eqref{eq:mass-terms}. Following the same deconstruction procedure as in \cite{deRham:2013awa}, we consider five-dimensional gravity and discretize the vielbein in the extra dimension. We denote by $x^\mu$ the continuous $3+1$ coordinates  and by $y$ the coordinate along the discretized extra dimension.
We perform a spatial ADM decomposition setting the lapse to unity and the shift to zero
\ba
\label{DvNgauge}
g^{(5d)}_{AB}(x,y)\d x^A \d x^B=\d y^2 + g\mn(x,y) \d x^\mu \d x^\nu=\d y^2+e_\mu^a(x,y)e_\nu^b(x,y)\eta_{ab}\d x^\mu \d x^\nu\,.
\ea
For simplicity we consider a discrete extra dimensions with two sites, localized respectively at $y=y_{1,2}$. We denote by $e$ the vielbein at the site $1$ and by $f$ the vielbein at the site 2, $e^a_\mu\equiv e_{\mu}^{a}(y_1)$ and $f^a_\mu\equiv e_{\mu}^{a}(y_2)$.
The derivative of  the vielbein on one site is then given
\ba
\partial_y e_{\mu}^{a}(y_1) = - m (e_{\mu}^{a}- f_{\mu}^{a})\,,
\ea
where the scale $m$ is related to the discretization scale (\ie the inverse distance between the sites).
In terms of the metric, this implies
\ba
K_{\mu\nu}(x^\mu, y_1)=\frac 12 \p_y g\mn(x,y_1)= - m \left( g_{\mu \nu}(x,y_1) - \frac{1}{2} (e_{\mu}^{a}  f_{\nu}^{b} + e_{\nu}^{a}  f_{\mu}^{b} )\right)\,.
\ea
Even for a fixed reference metric $f_{\mu\nu}=f_{\mu}^{a}f_{\nu}^{b}\eta_{ab}$, the vielbein formalism introduces an additional local Lorentz symmetry. We may use this symmetry to go to the Deser-van Nieuvenhuizen gauge $e_{\mu}^{a}  f_{\nu}^{b} \eta_{ab }= e_{\nu}^{a}  f_{\mu}^{b} \eta_{ab}$ for which $K_{\mu\nu}  = -m \left ( g_{\mu \nu} - e_{\mu}^{a}  f_{\nu}^{b} \eta_{ab} \right)= -m \left ( g_{\mu \nu} - g_{\mu \alpha}e^{\alpha}_{a}  f_{\nu}^{a}  \right)$
\cite{Deser:1974cy, Hoek:1982za}.
It is easy to show that as a consequence of this gauge choice,
\be
e^{\mu}_{a}  f_{\alpha}^{a} e^{\alpha}_{b}  f_{\nu}^{b} = e^{\mu}_{a}  f_{\alpha,b} e^{\alpha,a}  f_{\nu}^{b} =g^{\mu \alpha} f\mn\,,
\ee
and thus we find the following expression for the discretized extrinsic curvature
\be
\label{Ksqrt}
K^{\mu}_{\nu}(x^\mu, y_1)  =- m \left( \delta^\mu_{\nu} -\sqrt{g^{\mu \alpha} f_{\alpha \nu}  } \right) \equiv -m \K\mupn(g,f)\,.
\ee
Thus we see that the square root structure of $\K\mupn$ characteristic of the dRGT model of massive gravity follows automatically from discretizing the extra dimension directly in the vielbein language . \\

A specific example of dRGT massive gravity is obtained by taking the spatial ADM form for the action for five-dimensional gravity,
\ba
S^{5d}=\frac{M_5^3}{2}\int \d^5x \sqrt{-g^{(5d)}} R[g^{(5d)}]=\frac{M_5^3}{2}\int \d y \d^4 x \sqrt{-g}\(^{(4)}R[g]+[K]^2-[K^2]\)\,,
\ea
where we use the notation that square brackets represent the trace of a tensor and substituting in the discretized expression for the extrinsic curvature.

Now as explained in \cite{deRham:2013awa}, discretizing the extra dimension in the vielbein language is equivalent to replacing the extrinsic curvature with the above square root function \eqref{Ksqrt} of the metric and the reference metric and simultaneously replacing the integral over the extra dimension its projection over one site\footnote{Alternatively
one can also consider the sum of the different sites, $\int \d y  \L(x,y)\longrightarrow m^{-1} \sum_{i}\L(x,y_i)$, and obtain instead a theory of multi-gravity with as many interacting and dynamical spin-2 fields as there are sites.},
\ba
\label{replacement1}
\int \d y  \L(x,y)\hspace{10pt}&\longrightarrow& \hspace{10pt}m^{-1}\, \L(x,y_1)\\
K\mupn \hspace{10pt}&\longrightarrow& \hspace{10pt} - m \,\K\mupn(g,f)\,.
\label{replacement2}
\ea
In the case of two sites, this leads to a specific four-dimensional theory of massive gravity,
\ba
S^{4d}=\frac{\mpl^2}{2}\int \d^4 x \sqrt{-g}\(^{(4)}R[g]+m^2\([\K]^2-[\K^2]\)\)\,,
\ea
with
\ba
\mpl^2=\frac{M_5^3}{m}\,.
\ea
Moreover by changing the discretization every so slightly, \ie  for each of the two $y$-derivatives in $K^2$ one can give a different weight to the different sites,
\ba
(\partial_y e_{\mu}^a)(\p_y e^b_{\nu}) =4 m^2 (r e_{\mu}^{a}-(1-r) f_{\mu}^{a})(s e_{\nu}^{b}-(1-s) f_{\nu}^{b})\,,
\ea
with $0<r,s<1$ and we easily generalize the deconstruction procedure to obtain all the possible four-dimensional mass terms presented in section \ref{sec:NonLinearMassTerms}
\ba
S^{4d}=\frac{\mpl^2}{2}\int \d^4 x \(\sqrt{-g}\, ^{(4)}R[g]+\frac{m^2}{2}\(\Lmb_2+\alpha_3 \Lmb_3+\alpha_4 \Lmb_4\)\)\,,
\ea
with $\Lmb_j$ given fully non-linearly by eqn.~\ref{eq:mass-terms}, and $\alpha_{3,4}$ two dimensionless  constants related to $r$ and $s$.

We now follow the same procedure to  include the higher-dimensional Lovelock invariants and motivate a non-linear completion for new kinetic interactions for the graviton.

\subsection{Lovelock Interactions}

The deconstruction framework is easily generalizable to many sites and many extra dimensions. This is discussed in more detail in \cite{deRham:2013awa}. With these extra dimensions, come new Lovelock invariants which are a generalization of the scalar curvature which keep the equations of motion second order in derivatives. In $D=2n$ and $D=2n-1$ dimensions, there are $n$ such Lovelock invariants, which are given by
\ba
S_{\rm Lovelock}^{(j)}=\int \d^D x\ \sqrt{-g} &&\mathcal{E}^{\mu_1 \cdots \mu_{2j} \mu_{2j+1}\cdots \mu_D} \mathcal{E}^{\nu_1 \cdots \nu_{2j}}_{\ \ \ \ \ \ \ \mu_{2j+1}\cdots \mu_D} \nn \\
&&\times R_{\mu_1 \nu_1 \mu_2 \nu_2} \cdots R_{\mu_{2j-1}\nu_{2j-1}\mu_{2j}\nu_{2j}} \,,
\ea
for $j=1,\cdots,n=[(D+1)/2]$. The Lovelock invariant with $j=0$ is simply the cosmological constant, and $j=1$ corresponds to the well-known Einstein-Hilbert Ricci Scalar action. The Lovelock invariant which is quadratic in the curvature corresponds to the Gauss-Bonnet term,  which is a special combination of the Riemann tensor, the Ricci tensor and the scalar curvature,
\ba
\label{Lovelock}
S_{\rm Lovelock}^{(2)}= 4(D-5)! \int \d^D x \sqrt{-g}\(R_{\mu\nu\rho\sigma}^2-4 R_{\mu\nu}^2+R^2\)\,.
\ea
In four dimensions, this combination is a total derivative, but in dimensions larger than four, the Gauss-Bonnet term is dynamical. The same remains true for all the Lovelock invariants $j$ which are total derivatives in $D=2j$ dimensions and dynamical in dimensions larger than $2j$. When considering the deconstruction framework, it does therefore make sense to start  with the most general higher dimensional action which is free from any ghost-like pathology, namely to start in $D$ dimensions with all the $n$ Lovelock invariants and discretize the extra dimension(s).\\

The cosmological constant remains a cosmological constant in the lower dimensional picture after discretization. As seen previously the Einstein-Hilbert action leads to massive gravity in the lower-dimensional picture, and as we shall see below, the higher-order Lovelock invariants lead to kinetic interactions in the lower-dimensional picture.

\subsection{Deconstruction of Gauss-Bonnet}
In what follows we simply apply the deconstruction procedure presented in section \ref{sec:deconstructing-eh} and the replacements \eqref{replacement1}, \eqref{replacement2} to `deconstruct' five-dimensional Einstein-Gauss-Bonnet gravity. For that we perform a $4+1$ spatial ADM split and work in the five-dimensional Deser-van Nieuvenhuizen gauge \eqref{DvNgauge}.

As seen previously, the Einstein-Hilbert term leads to a specific four-dimensional theory of massive gravity (and the other possible mass terms can be obtained via alternative discretizations with different weight on the different sites). So in what follows we focus on the Gauss-Bonnet term. In the ADM decomposition, the Gauss-Bonnet Lagrangian \eqref{Lovelock} is given by
\ba
S_{\rm GB}&=& \frac{M_5^3}{m^2}\int \d^5 x \sqrt{-g}\({}^{(5)}R_{ABCD}^2-4 {}^{(5)}R_{AB}^2+{}^{(5)}R^2\)\nn \\
\label{GB_ADM}
&=& \frac{M_5^3}{4m^2} \int \d^5 x \sqrt{-g}\mathcal{E}^{\mu\nu\alpha\beta}\mathcal{E}^{\mu'\nu'\alpha'\beta'}
\Big[R_{\mu\nu\mu'\nu'}R_{\alpha\beta\alpha'\beta'}
-\frac{1}{12}K_{\mu\mu'}K_{\nu\nu'}K_{\alpha\alpha'}K_{\beta\beta'} \nn \\
&& \hspace{220pt}+K_{\mu\mu'}K_{\nu\nu'}R_{\alpha\beta\alpha'\beta'} \Big]\,,
\ea
where we emphasize that $\mathcal{E}^{\mu\nu\alpha\beta}$  represents the fully antisymmetric Levi-Cevita tensor.  We see appearing the four-dimensional Gauss-Bonnet contribution which is a total derivative in four dimensions,
\ba
\L_{\rm GB}^{4d}=\sqrt{-g}\mathcal{E}^{\mu\nu\alpha\beta}\mathcal{E}^{\mu'\nu'\alpha'\beta'}R_{\mu\nu\mu'\nu'} R_{\alpha\beta\alpha'\beta'}\,,
\ea
which can also be expressed in terms of the dual Riemann tensor
\ba
^* R^{\mu\nu\mu'\nu'}=\mathcal{E}^{\mu\nu\alpha\beta}\mathcal{E}^{\mu'\nu'\alpha'\beta'} R_{\alpha\beta\alpha'\beta'}\,,
\ea
which is transverse.

Upon substitution \eqref{replacement2}, we directly see appearing one of the mass terms combinations, $\L_4= \mathcal{E}\mathcal{E} \K \K \K \K$, which could also have been obtained from the Einstein-Hilbert curvature if one chose a non-trivial discretization, so the two first terms in \eqref{GB_ADM} are not fundamentally new. However the second line of \eqref{GB_ADM}  leads to a new non-trivial kinetic interaction which mixes both metric $\K\mupn$ to the Einstein tensor, leading to the new four-dimensional interaction
\ba
\label{KKR}
S^{4d}_{\K \K ^*R}= \mpl^2 \int\d^4x\sqrt{-g}\, \K_{\mu\nu}\K_{\alpha \beta}\ {}^*R^{\mu\alpha\nu\beta}\,.
\ea

As in the case for the mass terms, a different choice of discretization naturally leads to yet another interaction which can be obtained by simply performing the shift $\K\mn \to g\mn + \K\mn$.  We then immediately see that the new kinetic interaction \eqref{KKR} generalizes to yet an additional one (still in four dimensions)
\ba
\label{KG}
S^{4d}_{\K G}&=&-\frac{\mpl^2}{4} \int\d^4x\sqrt{-g}\, g_{\mu\nu}\K_{\alpha \beta}\ {}^*R^{\mu\alpha\nu\beta}\\
&=&  \mpl^2 \int\d^4x\sqrt{-g}\, \K\mn\ G^{\mu\nu}\,,
\ea
where $G_{\mu\nu}$ is the Einstein tensor.

As described in \cite{deRham:2013awa}, by changing the discretization procedure we will simply change the weights of the terms $S^{4d}_{\K G}$ and $S^{4d}_{\K \K ^*\!R}$.

Interestingly the interaction \eqref{KG} reduces to $\Lk_2$ at leading order (at that order \eqref{KG} is indistinguishable from the Einstein-Hilbert term) and the interaction \eqref{KKR} reduces to $\Lk_3$  also at its leading order (cubic order in that case). It is thus natural to expect that \eqref{KG} could be an alternative non-linear candidate for $\Lkb_2$ different than the Einstein-Hilbert term and \eqref{KKR} could be a completion for $\Lkb_3$. Unfortunately as we shall see below both these terms $\mathcal{L}^{4d}_{\K G}$ and $\mathcal{L}^{4d}_{\K \K ^*\!R}$ have an Ostrogradsky ghost and are thus not appropriate candidate for these completions.

\subsection{Ghosts in the Minisuperspace}

It is straightforward to see that these terms contain ghosts already in the minisuperspace approximation (which is a consistent truncation of the action)
\be
g_{00} = -N^2(t), \ \ \ g_{0i}=0,\ \ \ g_{ij} = a^2(t) \delta_{ij}.
\ee
Here $N(t)$ represents the lapse and the Hamiltonian ought to be linear in it.

In the minisuperspace approximation the two kinetic interactions become
\ba
S_{\K G} &=& 3 \int \d t \d ^3 x a^3 N \left(2\frac{\dot{a}^2}{a^2 N^2} - \frac{\dot{a}^2}{a^3 N^2}+\frac{\dot{a}^2}{a^2 N^3}\right) \nn \\
S_{\K \K ^* R} &=& 24 \int \d t \d ^3 x a^3 N \left(\frac{\dot{a}^2}{a^2 N^2} - \frac{\dot{a}^2}{a^3 N^2}+\frac{\dot{a}^2}{a^2 N^3} - \frac{\dot{a}^2}{a^3 N^3} \right).
\ea
Ghosts arise in both of these actions because of the terms scaling as $1/N^2$, which causes $N$ to appear non-linearly in the Hamiltonian form, violating \ref{Prop ADM}. Furthermore, it is clear that we cannot take any combination of the two actions to remove the ghost.

One possible out is to return to the original discretization in the vielbein. In order to express the kinetic terms in a metric language, we have assumed the DvN gauge condition \ref{DvNgauge} holds, which follows from discretizing the equation $\Om^{AB}_y=0$ for the spin connection in second order form. It is also possible to discretize in the vielbein in a first order form, where the vielbein $E^A$ and the spin connection $\Om^{AB}$ are treated independently, and we only solve for the spin connection after discretizing. Such a discretization procedure will introduce torsion, and there may not be a metric formulation. Exploring this possibility is beyond the scope of this work.

 Faced with this result we now open the spectrum of possibilities by systematically considering every possible term up to quartic order and show that only the Einstein-Hilbert term is allowed. The result holds beyond the quartic order expansion as we shall explain below.

%
%
%%%% Section 4: No Go Theorem
%
%
\section{No go theorem}
\label{sec:no-go}
The fact that the terms that arise naturally from deconstruction fail to propagate five degrees of freedom is already an indication that it may be impossible to find new kinetic interactions. Nevertheless, in this section we present an all orders perturbative proof that there can be no new kinetic interactions in massive gravity in four dimensions, so long as there is a local metric formulation of the interactions. \\

Let us first outline the argument before discussing the details:
\begin{itemize}
\item We will write down the most general Lorentz invariant Lagrangian of the form $(\pa H)^2 H^n$, up to quartic order in $H$. There are a total of $4+14+38=56$ parameters.
\item By demanding that the Lagrangian satisfied \ref{Prop DL}, we will be left with a 4 parameter family of potentially BD-ghost-free Lagrangians in four space-time dimensions.
\item We then perform a perturbative ADM analysis on the remaining 4 parameter Lagrangian and find that there is only one choice of parameters that does respect \ref{Prop ADM}. The resulting uniquely determined Lagrangian is equivalent to the Einstein-Hilbert action at quartic order. This means that only the Einstein-Hilbert term is an allowed kinetic term, and the derivative term $\Lk_3$ that started at cubic order has no completion.
% \ba
%\begin{array}{ccc}
%\text{Leading order}\hspace{20pt} & &\hspace{30pt} \text{Fully non-linear completion}\\[5pt]
%\Lk_2 & \hspace{30pt}\longrightarrow\hspace{30pt}&\hspace{30pt} \Lkb_2 =\sqrt{-g}\ R[g]  \\
%&\hspace{30pt}\longrightarrow\hspace{30pt}&\hspace{30pt} \text{No other completions.}  \\[3pt]
%\Lk_3 &\hspace{30pt} \longrightarrow\hspace{30pt}&\hspace{30pt} \Lkb_3 \text{does not exist.}
%\end{array}\nn
%\ea
\ba
\begin{array}{ccc}
\text{Leading order}\hspace{20pt} & &\hspace{30pt} \text{Fully non-linear completion}\\[3pt]
\hline \\
\Lk_2 & \hspace{30pt}\longrightarrow\hspace{30pt}&\hspace{30pt}
\begin{cases}
 \Lkb_2 =\sqrt{-g}\ R[g]  \\
  \text{No other completions.}
  \end{cases} \\[5pt]
\Lk_3 &\hspace{30pt} \longrightarrow\hspace{30pt}&\hspace{30pt} \Lkb_3 \text{does not exist.}
\end{array}
\ea
\item If there were any new allowed kinetic term it should satisfy \ref{Prop lin Stuc}. But as we have seen in section \ref{sec:NewMassTerms}, only $\Lk_2$ and $\Lk_3$ satisfy \ref{Prop lin Stuc} so there can be no new kinetic term that arises beyond the cubic order. Since $\Lk_3$ has no completion and $\Lk_2$ has a unique completion as the Einstein-Hilbert term, this is the only possible kinetic term for massive gravity. This completes the all orders argument.
\end{itemize}

\subsection{Decoupling limit}
\label{sec:decoupling-limit}
As a first step we will consider the decoupling limit of the possible kinetic interactions. In the massive gravity limit we can write a general theory with kinetic interactions as
\be
S= \frac{\mpl^2}{2}\int \d ^D x \sqrt{-g} R + \frac{m^2 \mpl^2}{2} \int \d^D x  \sqrt{-g} \sum_n \alpha_n \Lmb_n+ \Lambda_{\rm der}^2 \int \d^D x  \mathcal{L}_{\rm der}\,,
\ee
where $\mathcal{L}_{\rm der}$ is fully non-linear and should include the non-linear completion of $\Lk_3$ if it exists and any other non-linear completion of $\Lk_2$.
We consider Lagrangians of the schematic form $\mathcal{L}_{\rm der} \sim(\partial^2 H^n)$, where $H$ is the unitary gauge metric fluctuation.\\

We have a scale $\Lambda_{\rm der}$ that is not fixed in terms of the other parameters. The only requirement is that one cannot scale the theory in such a way that $\mathcal{L}^{(m)}_n+ \Lk$ remains without $\Lmb_2$, (as seen earlier, new kinetic terms can only be present if the graviton is massive).
In order to be able to perform the standard decoupling limit scaling, we take $\Lambda_{\rm der}^2=\mpl \Lambda_3 = \Lambda_3^4 / m^2$, although the results are independent of that very scale. If we change the scale, $\Lambda_{\rm der}$ we could still consider an equivalent decoupling and keep track of only the terms at most linear in $h$ or quadratic in $A$, while keeping the terms to all order in what is then the helicity-0 mode $\pi$.

As discussed in Section \ref{sec:linear-diff}, the new interaction in \cite{Hinterbichler:2013eza} was found using a linear \stu decomposition. That is sufficient to find the leading order piece of an interaction. However to find a non-linear completion we will need to use the nonlinear \stu decomposition \eqref{eq:non-linear-stuck}.
%\begin{equation}
%\label{eq:nonlinear-stuck}
%H_{\mu\nu}=\hat{h}_{\mu\nu}+ \partial_{(\mu}\hat{A}_{\nu)} + 2\hat{\Pi}_{\mu\nu}+\eta^{\rho \sigma} \left(\partial_\mu \hat{A}_\sigma +  \hat{\Pi}_{\mu\sigma}\right)\left(\partial_\rho \hat{A}_\nu +\hat{\Pi}_{\rho \nu}\right).
%\end{equation}
%with $\hat{\Pi}_{\mu\nu}\equiv \partial_\mu\partial_\nu \hat{\pi}$. The hatted dimensionless fields can be expressed in terms of the canonically normalized fields by
%\ba
%\hat{h}_{\mu\nu}=\frac{h_{\mu\nu}}{\mpl},\ \ \hat{A}_\mu=\frac{A_\mu}{(m \mpl)^{1/2}},\ \ \hat{\pi}=\frac{\pi}{(m^2 \mpl)^{1/3}}.
%\ea

We would like to ensure that there are only five propagating degrees of freedom. So long as the interaction has 1 or fewer powers of $h$, this is equivalent to the condition that the equations of motion for all fields have two or fewer derivatives. However once we consider interactions with higher powers of $h$, the analysis becomes more subtle. The interpretation of $A_\mu$ and $\pi$ as the physical helicity 1 and 0 modes, which works around flat space, does not work around a general curved background because the curvature introduces mixings between the fields. At linear order in $h$ this poses no problem, because we may think of the background metric as being $\eta$. However at higher order in $h$ we must think of $\eta+h$ as the background metric. Stated another way, when we move to higher orders in $h$ we should really introduce the \stu fields as four diff scalars $\phi^a$. In order to propagate five degrees of freedom, we must check that the hessian $\pa^2 \mathcal{L}_{\rm der} / \pa \dot{\phi^a} \pa \dot{\phi^b}$ has a zero eigenvalue. In the decoupling limit, we may simply interpret this as saying that $\pi$ must have second order equations of motion. However in general beyond the decoupling limit (\ie at higher order in $h$), this subtlety becomes important. If one naively computes the equations of motion for $\phi^0$, one would find higher order equations of motion, but a more careful analysis shows that there is one constraint among the equations of motion for the $\phi^a$ and so only $3\times 2$ pieces of initial data are needed to specify the time evolution of the four \stu $\phi^a$. For more details see \cite{deRham:2011rn}.  \\

As a result, when writing down terms in the action beyond linear order in $h$, we may not conclude that a ghost is present simply by finding that the equations of motion include third or higher time derivatives acting on $h$, $A$, and $\pi$, since these are not the true physical degrees of freedom. Thus we will only work to linear order in $h$ in the decoupling limit. This is the essence of \ref{Prop DL}.
As we will see even this requirement alone is quite constraining. We will then further constrains the terms by performing a genuine ADM analysis which is valid beyond the decoupling limit. This corresponds to \ref{Prop ADM}.\\

In the decoupling limit, a general interaction of the form $\pa^2 H^n$ looks like
\ba
\mathcal{L}_{\rm der}^{(n)} &\sim &\frac{\Lambda_3^4}{m^2} H^{n-2} \partial H \partial H \nn \\
& \sim &  m^{2n_h + n_A - 2} \Lambda_3^{4-3(n_h + n_A + n_\pi)} (\pa^2) h^{n_h} (\pa A)^{n_A} (\pa^2 \pi)^{n_\pi}.
\ea
Since the \stu decomposition is nonlinear, in general $n\ne n_h + n_A + n_\pi$. By inspecting this equation, we see that we may use our requirement that the equations of motion be at most second order in derivatives for all interactions up to the scale $\Lambda_3$ and still have $n_h \leq 1$.\\

Once we eliminate the dangerous interactions at this scale, we can no longer use the criteria of second order equations of motion to check for ghosts. Any higher scale involves terms with $n_h >1$. \footnote{Note that this analysis does not crucially depend on choosing $\Lambda_{\rm der}^2 = \Lambda_3^4/m^2$. Keeping a general $\Lambda_{\rm der}$, we still see that canceling interactions coming in at a scale below any interaction with $n_h >1$ amounts to canceling all interactions with general $n_\pi$ and either $n_A \leq 2, n_h =0$ or $n_A =0, n_h \leq 1$. Thus it is not possible to avoid this decoupling limit analysis by simply choosing a smaller value for $\Lambda_{\rm der}$. However the choice $\Lambda_{\rm der}^2=\Lambda_3^4/m^2$ allows for a nicer physical interpretation of the result and so we will make this choice. } \\

Note that unlike the case for the mass terms, all interactions which come in at the scale $\Lambda_3$ or below will have too many derivatives per field to have second order equations of motion. This means we must cancel \emph{all} interactions (up to total derivatives) that come in at $\Lambda_3$.

\subsubsection{Relationship with standard \stu analysis}

In this work we are not working with the \stu analysis in its usual representation, but rather using the non-linear helicity decomposition, as described in \cite{deRham:2011qq}. In particular we are not using attempting to reintroduce covariance. Rather, we are remaining in unitary gauge, and simply performing a field redefinition in order to identify the physical degrees of freedom.

We may at any stage move to the standard \stu language by performing the gauge transformation to the metric $g^S_{\mu\nu}$ as follows

\begin{equation}
g_{\mu\nu} (x^a) = g^{S}_{ab} (\Phi^a) \partial_\mu \Phi^{a} \partial_\nu \Phi^{b},
\end{equation}
with $\Phi^a = x^a + \frac{1}{m M_{\rm Pl}}A^a + \frac{1}{m^2 M_{\rm Pl}}\partial^a \pi$. If we further decompose $g^S_{\mu\nu}=\eta_{\mu\nu}+\frac{1}{M_{\rm Pl}}h^S_{\mu\nu}$, then in this gauge, $H_{\mu\nu}$ is given by
\ba
H_{\mu\nu} &=& g_{\mu\nu} - \eta_{\mu\nu} \nn \\
&=&\frac{1}{M_{\rm Pl}} h^S_{ab}(\Phi^a) \partial_{\mu} \Phi^a \partial_\nu \Phi^b + \frac{1}{m \mpl}\partial_{(\mu}A_{\nu)} + \frac{2}{m^2 \mpl} \pa_\mu \pa_\nu \pi \nn \\
&&-\eta^{\rho \sigma} \left( \frac{\partial_{\mu} A_{\rho}}{m \mpl}+  \frac{\pa_\mu \pa_\rho \pi}{m^2 \mpl}\right)\left( \frac{\partial_{\sigma} A_{\nu}}{m \mpl}+  \frac{\pa_\sigma \pa_\nu \pi}{m^2 \mpl}\right).
\ea

Equating this with our field redefinition \ref{eq:non-linear-stuck}, we see that the unitary gauge helicity-2 field $h_{\mu\nu}$ that we work with in this paper is related to the helicity-2 field in the \stu gauge by
\begin{equation}
h_{\mu\nu} = h^S_{ab}(\Phi^a) \partial_{\mu} \Phi^a \partial_\nu \Phi^b.
\end{equation}

The BD-ghost-free analysis of this section may be performed in either gauge. However we can make a stronger statement: the analysis of the kinetic interactions is the same in both languages in the decoupling limit. To see this, first note that in the decoupling limit, at linear order in the unitary gauge helicity 2 field $h_{\mu\nu}$ we can write the derivative interactions as
\begin{equation}
\mathcal{L}_{\rm der} = h_{\mu\nu} Y^{\mu\nu},
\end{equation}
where $Y^{\mu\nu} \sim \sum_{n_\pi} \Lambda_3^{4-3n_\pi} \partial^2 \left(\partial^2 \pi \right)^{n_\pi}$. Since $Y^{\mu\nu}$ is higher derivative in $\pi$, we must have that $Y^{\mu\nu}=0$.

We may then restate this analysis in terms of the \stu gauge helicity 2 field $h^S_{\mu\nu}$ as follows
\ba
\mathcal{L}_{\rm der} &=& h^S_{ab}(\Phi^a) \partial_{\mu} \Phi^a \partial_\nu \Phi^b Y^{\mu\nu} \nn \\
&=& h^S_{\mu\nu}(x)\left(Y^{\mu\nu} + \frac{1}{\Lambda_3^3} \left(2 \partial_\sigma \partial^\mu \pi Y^{\sigma \nu} - \partial_\sigma (Y^{\mu\nu}) \partial^\sigma \pi \right)+ \cdots \right) \nn \\
&=& h^S_{\mu\nu}(x) Y^{S,\, \mu\nu},
\ea
where in the last line we have defined $Y^S = Y + \frac{1}{\Lambda_3^3} \partial(\partial \pi Y) + \cdots$. Since the relationship between $Y$ and $Y^S$ is trivially invertible, and since we must have that $Y=0$, it follows that $Y^S=0$ as well. Thus we see that the same condition must be imposed in both gauges, namely that all terms with $n_h=1$ must vanish. Furthermore since the fields $\pi$ and $A_\mu$ are the same in both gauges, as discussed in \cite{deRham:2011qq}, the interactions with $n_h=0$ must vanish in the decoupling limit in both gauges as well. Thus while we will continue to work in the unitary gauge description, the conclusions of this section would be unchanged had we done the analysis in the \stu gauge.

\subsubsection{Quadratic Order}

At quadratic order the most general term of the form $\pa H \pa H$ has 4 free parameters. However, since any interaction must satisfy \ref{Prop lin Stuc} at leading order, there is only one possible term at quadratic order, which is nothing else but the linearized Einstein-Hilbert term denoted by $\mathcal{L}_a^{(2)}$ for reasons that will be clear later
\be
a \mathcal{L}^{(2)}_{a}= -2a \Lk_2
% H^{\mu\nu}\left(\mathcal{E} H\right)_{\mu\nu} = a H^{\mu\nu}\left(\square H_{\mu\nu}  + \partial_\alpha \partial_{(\mu} H^\alpha_{\ \ \nu)}+\partial_\mu \partial_\nu H + \eta_{\mu\nu}\left(\partial_\alpha \partial_\beta H^{\alpha\beta}-\square H\right)\right).
\ee
Note once again that only the helicity-2 part of the spin-2 field enter that kinetic term (even though this was not a requirement) and the helicity-1 and -0 should thus gain their kinetic term from $\Lm_2$.  The proper normalization for the helicity-2 mode sets $a=1/8$.

\subsubsection{Cubic Order}

Now we move onto the most general term at cubic order of the form $\partial^2 H^3$. We may use integration by parts to put the action in the form $H \partial H \partial H$, and we are left with 14 terms\footnote{There are 16 possible contractions of the form $H\partial H \partial H$, but 2 linear combinations of these are total derivatives.}
\ba
\mathcal{L}^{(3)}_{\rm gen} &=& b_1 H^{\mu\nu}\pa_\mu H^{\rho\sigma}\pa_\nu H_{\rho\sigma} + b_2  H^{\mu\nu} \pa_\mu H^\sigma_{\ \ \nu}\pa_\sigma H + b_3 H^{\mu\nu} \pa_\mu H \pa_\nu H   \nn \\
&+& b_4 H^{\mu\nu} \pa_\sigma H \pa^\sigma H_{\mu\nu} + b_5 H \pa_\mu H \pa^\mu H + b_6 H^{\mu\nu} \pa_\rho H_{\mu \sigma}\pa^\sigma H^\rho_{\ \ \nu} \nn \\
&+& b_7 H \pa_\mu H_{\nu\sigma} \pa^\sigma H^{\mu\nu} + b_8 H^{\mu\nu} \pa_\rho H_{\mu \sigma}\pa^\rho H^\sigma_{\ \ \mu} + b_9 H \pa_\mu H_{\nu \sigma} \pa^\mu H^{\nu\sigma} \nn \\
&+& b_{10} H^{\mu\nu} \pa_\sigma H^\sigma_{\ \ \mu} \pa_\rho H^\rho_{\ \ \nu} + b_{11} H^{\mu\nu} \pa_\mu H^\rho_{\ \ \nu} \pa_\sigma H^\sigma_{\ \ \rho} + b_{12} H \pa_\mu H^{\mu\nu}\pa_\sigma H^\sigma_{\ \ \nu} \nn \\
&+& b_{13} H^{\mu\nu} \pa^\rho H_{\mu\nu}\pa^\sigma H_{\rho \sigma}+ b_{14} H \pa^\mu H \pa^\nu H_{\mu\nu}.
\ea
We take the total Lagrangian up to cubic order, including the ghost free contribution from quadratic order
\be
\mathcal{L} = a \mathcal{L}^{(2)}_{a} + \mathcal{L}^{(3)}_{\rm gen}
\ee
and perform the nonlinear \stu decomposition given by Eq.~\ref{eq:non-linear-stuck}. Note that it is crucial to keep $\mathcal{L}^{(2)}_{a}$ because of the non-linearity of the \stu decomposition. \\

We then focus on interactions that come in at or below the scale $\Lambda_3$ and demand that the equations of motion contain fewer than two time derivatives per field. This results in 12 independent conditions on the 14 parameters cubic order parameters $b_j$. The result is that there are three free parameters, which we may take to be $a, b_1, b_2$. We emphasize that the nonlinear \stu decomposition is crucial, this fixes some of the $b_j$ in terms of $a$ (as is already expected from General Relativity).
We can then write the most general Lagrangian, which is ghost free up to the scale $\Lambda_3$ (\ie that satisfy \ref{Prop DL}), as
\ba
\mathcal{L}_{{\rm g.f.},\Lambda_3}^{(3)} &=& a \left( \mathcal{L}^{(2)}_a + \mathcal{L}^{(3)}_a\right)+ b_1 \mathcal{L}^{(3)}_{b_1} + b_2 \mathcal{L}^{(3)}_{b_2} \nn \\
&=& A_{EH} \mathcal{L}_{EH} + A_{\K G} \mathcal{L}_{\K G} + A_{\K \K ^*\! R} \mathcal{L}_{\K \K ^*\! R},
\ea
where $\mathcal{L}^{(3)}_{a, b_{1,2}}$ are defined in (\ref{L3a}-\ref{L3b2}) and
\be
\label{eq:A_coefficients}
A_{EH}= 4 a + 32 b_1 - 8 b_2, \ \ A_{\K G} = 32 b_1 - 8 b_2, \ \ A_{\K\K^*\!R} = a - 3 b_1 + b_2,
\ee
and where the nonlinear terms may be expressed up to cubic order as
\ba
 \mathcal{L}_{EH} &=&  \sqrt{-g} R = \frac{1}{4}\left[\mathcal{L}_a - \mathcal{L}_{b_1} - 4 \mathcal{L}_{b_3} \right]\nn \\
 \mathcal{L}_{\K G} &=& \sqrt{-g} \K_{\mu\nu}G^{\mu\nu} = -\frac{1}{4}\left[ \mathcal{L}_a - \frac{3}{2} \mathcal{L}_{b_1} - \frac{11}{2} \mathcal{L}_{b_2}\right] \nn \\
\mathcal{L}_{\K \K ^*\! R} &=& \sqrt{-g} \K_{\mu\nu}\K_{\mu' \nu'}\, ^*\!R^{\mu\mu'\nu\nu'}= \mathcal{L}_{b_1} + 4 \mathcal{L}_{b_2} = \Lk_3.
\ea
The explicit form of $\mathcal{L}_{{\rm g.f.},\Lambda_3}$ is given in \ref{eq:L3Lambda3}. Note that expressing $\mathcal{L}_{{\rm g.f.},\Lambda_3}$ in terms of cubic expansions of the non-linear terms is merely meant to be suggestive. Of course any nonlinear term expressible in terms of the parameters $a$, $b_1$, and $b_3$ at cubic order would be ghost free at this level as well. \\

Interestingly, the interactions arising from deconstruction are free of ghosts up to the scale $\Lambda_3$. The term $\mathcal{L}_{\K\K^*\!R}$ is manifestly equivalent to $\Lk_3$ at cubic order. More interesting is the term $\mathcal{L}_{\K G}$, which is degenerate at quadratic order with the Einstein-Hilbert term, but becomes different at cubic order. Of course, this term has a ghost at higher scales as we have seen in Section \ref{sec:deconstruction}. Nevertheless, deconstruction generates a natural guess for the kinetic interactions, that pass a nontrivial check.\\

Thus up to cubic order we see two remarkable facts:
\begin{itemize}
\item First we re-confirm that the appearance of a new term $\Lk_3$, which is independent of $\Lk_2$, \ie $\Lk_3$ does not arise as the cubic extension of $\Lk_2$ but rather as its own new and independent kinetic interaction as was already argued in \eqref{sec:NewKineticTerms}.
\item  Second we see that $\Lk_2$ can have two different completions up to cubic order in the decoupling limit. The first one $\mathcal{L}_{EH}$ is nothing else but the Einstein-Hilbert term  to that order. This is exactly as expected from \eqref{kin extensions}. However there is another possible completion of the linearized Einstein-Hilbert term $\Lk_2$ at cubic order which is not the cubic Einstein-Hilbert term but rather the term $\mathcal{L}_{\K G}$. So this term could be a natural candidate for the second line of \eqref{kin extensions}, but as we have already seen this term has actually a ghost non-linearly and we will confirm this by acting \ref{Prop ADM}.
\end{itemize}

\subsubsection{Quartic Order}
At quartic order the analysis is very similar as that at cubic order. We start with the most general terms of the form $\partial^2 H^4$. After removing all total derivative combinations there are 38 free parameters:
\be
\mathcal{L}_4 = \sum_{j=1}^{38} c_j \left[H^2 \pa H \pa H\right]_j.
\ee
The 38 contractions $H^2 \pa H \pa H$ are written explicitly in \ref{eq:L4gen}. Actually in four and fewer dimensions one of these combinations cancels exactly, but we shall keep it for now as the analysis is applicable to any number of dimensions. \\

Repeating the procedure we followed at cubic order, we find the most general ghost free quartic lagrangian can be written in terms of five parameters: $a$, $b_1$, and $b_2$, as well as two new parameters $c_1$ and $c_2$
\ba
\mathcal{L}^{(4)}_{{\rm g.f.},\Lambda_3} &=& a \left( \mathcal{L}^{(2)}_a + \mathcal{L}^{(3)}_a + \mathcal{L}^{(4)}_a\right) + b_1 \left( \mathcal{L}^{(3)}_{b_1} + \mathcal{L}_{b_1}^{(4)} \right)+ b_2 \left(\mathcal{L}^{(3)}_{b_2} + \mathcal{L}^{(4)}_{b_2}\right) \\
&&+ c_1 \mathcal{L}^{(4)}_{c_1} + c_2 \mathcal{L}^{(4)}_{c_2}\,, \nn
\ea
where $\mathcal{L}^{(4)}_{c_{1,2}}$ are expressed in \eqref{L4c1} and \eqref{L4c2} and $\L^{(4)}_{a,b_{1,2}}$ in (\ref{L4a}-\ref{L4b1}).

$\mathcal{L}^{(4)}_{c_{1}}$ and $\mathcal{L}^{(4)}_{c_{2}}$ represent interactions appearing at the quartic level that are not completions to the previous cubic or quadratic terms (they enter independently from $\mathcal{L}_a$ and from $\mathcal{L}_{b_{1,2}}$). Thus these are `new' kinetic interactions. However we know that in four dimensions there can be no new  terms that satisfy \ref{Prop lin Stuc} so these terms should not be present in four dimensions. The resolution lies in the following two facts:

First, at the moment we have only applied the \ref{Prop DL}, and not yet \ref{Prop lin Stuc} nor \ref{Prop ADM}. As we shall see below \ref{Prop ADM} removes one of these new kinetic quartic terms (and so would \ref{Prop lin Stuc}).

Second,
so far this analysis has worked for general space-time dimension $D$. If we now specialize to $D=4$, the following combination is identically zero,
\be
\mathcal{L}^{(4)}_{c_1} + 4 \mathcal{L}^{(4)}_{c_2} \equiv 0.
\ee
In $D>4$, this term corresponds to a term invariant satisfying \ref{Prop lin Stuc} at its leading quartic order
and can be written as
\be
\mathcal{L}_{c_1} + 4 \mathcal{L}_{c_2} = \frac{1}{12  (D-5)!} \epsilon^{a_1 \cdots a_D} \epsilon^{b_1 \cdots b_D}  \left(\pa_{a_1} H_{a_2 b_2} \right)\left(\pa _{b_1} H_{a_3 b_3} \right) H_{a_4 b_4} H_{a_5 b_5} \prod_{5 < j \leq D}\eta_{a_j b_j}.
\ee
Then redefining $c_1 - \frac{c_2}{4} \rightarrow c_1$ the Lagrangian has 4 parameters
\ba
\label{eq:quarticOrderDL}
\mathcal{L}^{(4)}_{{\rm g.f.},\Lambda_3} &=& a \left( \mathcal{L}^{(2)}_a + \mathcal{L}^{(3)}_a + \mathcal{L}^{(4)}_a\right) + b_1 \left( \mathcal{L}^{(3)}_{b_1} + \mathcal{L}_{b_1}^{(4)} \right)+ b_2 \left(\mathcal{L}^{(3)}_{b_2} + \mathcal{L}^{(4)}_{b_2}\right) + c_1 \mathcal{L}^{(4)}_{c_1} \hspace{20pt} \\
&=& a \mathcal{L}_a + b_1 \mathcal{L}_{b_1} + b_2 \mathcal{L}_{b_2} + c_1 \mathcal{L}_{c_1}\nn \\
&=& A_{EH} \mathcal{L}_{EH} + A_{\K G} \mathcal{L}_{\K G} + A_{\K \K ^*\! R} \mathcal{L}_{\K \K ^*\! R} + A_4 \mathcal{L}_{c_1}\,, \nn
\ea
where the non-linear Lagrangians $\mathcal{L}_{a,b_1,b_2,c_1}$ are defined in (\ref{La}-\ref{Lc1}) and $A_{EH},A_{\K G}, A_{\K \K ^*\! R}$ are given by \eqref{eq:A_coefficients} with $A_{4} = \frac{a}{8} - \frac{63}{8} b_1 + 2 b_2 + c_1$.

The Einstein-Hilbert and the terms arising from deconstruction may be written in terms of these new parameters as
\ba
 \mathcal{L}_{EH}&=& \frac{1}{4}\left[ \mathcal{L}_a - \mathcal{L}_{b_1} - 4 \mathcal{L}_{b_2} \right]   \nn \\
\mathcal{L}_{\K G}&=&  -\frac{1}{4}\left[ \mathcal{L}_a - \frac{3}{2} \mathcal{L}_{b_1} - \frac{11}{2} \mathcal{L}_{b_2}- \frac{15}{16} \mathcal{L}_{c_1} \right] \nn \\
\mathcal{L}_{\K\K^*\!R}&=& \mathcal{L}_{b_1} + 4 \mathcal{L}_{b_2} - \frac{1}{8} \mathcal{L}_{c_1} .
\ea
The term $\mathcal{L}_{c_1}$ does not correspond to a non-linear completion found by deconstruction, nor satisfying \ref{Prop lin Stuc}, however we will show in the next section that it has indeed a ghost so we will not attempt to construct a candidate non-linear completion.

\subsection{Perturbative ADM analysis}
We will now perform a perturbative ADM analysis on the action written in unitary gauge.
We could have started by performing the ADM analysis straight away rather than first going to the decoupling limit. However the decoupling limit analysis performed previously is (a) more physical (it identifies the propagating degrees of freedom and the potential presence of Ostrogadsky ghosts), (b) it can be performed without needing to perform a $(3+1)$-split and is thus more efficient at eliminating sets of terms, (c) for the mass terms the decoupling limit analysis was sufficient to deduce the correct non-linear completion for all the mass terms.  Here we see that the last point does not apply for the kinetic terms, but \ref{Prop DL} was still sufficient to eliminate most of the possible kinetic interactions that arise up to quartic level. We now see that \ref{Prop ADM} eliminates the three last free coefficients (namely $b_1, b_2$ and $c$). \\

We start with the action in unitary gauge (\ie fix the gauge where $A^\mu=0$ and $\pi=0$) and work with the metric in ADM form
\be
g_{\mu\nu}\d x^\mu \d x^\nu = - N^2 \d t^2 + \gamma_{ij}\left( N^i \d t + \d x^i \right)\left( N^j \d t +  \d x^j \right).
\ee
We then expand $g_{\mu\nu}$ around flat space
\be
g_{\mu\nu} = \eta_{\mu\nu} + H_{\mu\nu},
\ee
and express the action $\mathcal{L}^{(4)}_{{\rm g.f.},\Lambda_3}$ in terms of the ADM variables so as to apply \ref{Prop ADM} on the quartic order Lagrangian \eqref{eq:quarticOrderDL}.

\subsubsection{The ultralocal limit}

First we work in the ultralocal limit. This will eliminate two of the four possible free parameters we found in the decoupling limit analysis. We expand the metric around around flat space\footnote{Technically this is not a consistent truncation, because we are introducing a preferred direction in $\delta N^i$ without maintaining the isotropy in $\gamma_{ij}$. However, our conclusions will not depend on the detailed form of the Hamiltonian, but merely on the existence of very dangerous terms at the Lagrangian level, and so we for the result we find the truncation we use is consistent.}
\ba
N &=& 1 + \delta N(t)  \nn \\
N^i &=&  \delta N^i (t) \nn \\
\gamma_{ij} &=& \left(1 + \delta a(t)\right) \delta_{ij}.
\ea
At cubic order, we find that there is a contribution at the level of the Lagrangian
\be
\mathcal{L}^{(3)}_{{\rm g.f.},\Lambda_3} \supset (4b_1 - b_2) \frac{\d (\delta N_i)^2}{\d t} \frac{\d (\delta a)}{\d t}.
\ee
This term violates \ref{Prop ADM} and cannot be put in first order form without involving time-derivatives of the shift. This means that if $4b_1-b_2\ne 0$ there would be new phase space variables which would signal the presence of an Ostrogadsky ghost\footnote{The primary and secondary Hamiltonian constraints present in massive gravity are sufficient to remove two of the phase variables in massive gravity. Furthermore it has been shown that there are no tertiary constraints in massive gravity \cite{Hassan:2011ea}, which means that there can be no tertiary constraints in this theory either. So if the shift were also part of the phase space variables, there could not be enough constraints to lead to only five physical degrees of freedom.}. Thus we impose the condition $4b_1 = b_2$. \\

After imposing this condition, at quartic order we find a similarly dangerous term
\be
\mathcal{L}^{(4)}_{{\rm g.f.},\Lambda_3} \supset c_1 \frac{\d (\delta N_i)^2}{\d t} \frac{\d (\delta a)^2}{\d t}.
\ee
This must be cancelled by setting $c_1=0$. Thus in $D\le4$ there can be no contribution from a term that starts at quartic order. This was already anticipated from \ref{Prop lin Stuc} but we have provided here a consistency check. \\

After imposing these two conditions we have a two-parameter family of quartic order Lagrangians
\be
\mathcal{L}^{(4)}_{\Lambda_3 + m.s.} = a \mathcal{L}_a + b_1 \left(\mathcal{L}_{b_1} + 4 \mathcal{L}_{b_2}\right)= 4 a \mathcal{L}_{EH} + 4 (a+b_1) \left(\mathcal{L}_{b_1} + 4 \mathcal{L}_{b_2}\right).
\ee
Interestingly, the extra parameter corresponds to the natural extension of $\Lk_3$ to quartic order.

\subsubsection{Including Inhomogeneities}

Finally we can eliminate the final parameter by allowing for some spatial dependence in the $x$ direction. We take
\ba
N &=& 1 + \delta N(t,x)  \nn \\
N^i &=&  \delta N^i (t,x) \nn \\
\gamma_{ij} &=& \left(1 + \delta a(t,x)\right) \delta_{ij}.
\ea
The Lagrangian contains the term
\be
(a+b_1) \frac{\d \delta N_i}{\d t} \delta N^i \delta N_j \pa_i \delta a.
\ee
This gives a kinetic term for $N_i$, adding 3 new phase space degrees of freedom to the system. These are 3 degrees of freedom in addition to the 3 degrees of freedom that arise from violating diffeomorphism invariance. Since the kinetic term for the $N_i$ degrees of freedom starts at cubic order, these extra degrees of freedom would be strongly coupled. It is also known that if the graviton propagates more than five degrees of freedom, the extra modes inevitably lead to instabilities \cite{Boulware:1973my}. \\

We remove this dangerous term by imposing $a=-b_1$. As a result, up to quartic order the unique kinetic interaction that does not propagate more than five degrees of freedom in four dimensions is simply the Einstein-Hilbert term
\be
\mathcal{L}_{{\rm g.f.},\Lambda_3+ADM}^{(4)} = 4 a\left[ \mathcal{L}_a - \left(\mathcal{L}_{b_1} + 4 \mathcal{L}_{b_2}\right) \right] = 4 a \mathcal{L}_{EH}.
\ee
In other words, any new kinetic interaction must vanish at quartic order or be exactly degenerate with the Einstein-Hilbert term up to quartic order.

\subsection{All orders no-go}
Consider a generic nonlinear kinetic interaction, that begins at $\mathcal{O}(H^n)$ with $n\ge 5$
\be
\Lkb_n(H_{\mu\nu}) = \Lambda_{\rm der}^2 \sum_{k=n}^\infty\beta_k \pa^2 H^k = \Lambda_{\rm der}^2 \(\Lk_n+\text{sub-leading terms}\)\,,
\ee
where all indices are raised and lowered with respect to the flat reference metric. We emphasize that in this notation the index $n$ denotes the order at which this new family of kinetic interactions starts. No matter at which order it starts, we expect an infinite number of subleading contributions to it, and so $\Lkb_n(H_{\mu\nu})$ is fully non-linear and contains terms of all order $k\ge n$. For the sake of the argument we separate out the leading piece $\Lk_n$ which is genuinely $n^{\rm th}$-order from the sub-leading contributions. For any  $\Lk_n$ we think of $\Lkb_n$ as its non-linear completion which satisfies \ref{Prop ADM}.\\

Generically this term will propagate more than five degrees of freedom.
As explained in section~\ref{sec:linear-diff}, in order for it to propagate five or fewer degrees of freedom, this terms needs to satisfy \ref{Prop lin Stuc}:  $\Lk_n$ cannot contain terms with more than two derivatives on each field at the level of the equation of motion when performing a {\it linear} \stu decomposition \eqref{eq:linear-stuck}.

The reason for this requirement was explained in section~\ref{sec:FPtheory}. To reiterate the essence of the argument, at leading order in a new set of interactions, all fields can be treated as if they were living on flat space and the linear \stu decomposition \eqref{eq:linear-stuck} correctly identifies the propagating degrees of freedom ($h\mn$ as the helicity-2 mode, $A_\mu$ as the helicity-1 and $\pi$ as the helicity-0 mode). It thus follows that if any of these fields admits equations of motion with more than two derivatives (which cannot be removed by substitution of the other field equations of motion), then there will be a genuine higher-order Ostrogadsky instability signaling the presence of a new degree of freedom in addition to the five expected ones, namely a BD ghost. \\

If $\Lk_n$ already satisfies \ref{Prop lin Stuc},  then we can ask the question of whether or not it admits a non-linear completion $\Lkb_n$.  In that case
the linear \stu decomposition should be replaced by the non-linear one. For instance if $\beta_{k+1}\pa^2 H^{k+1}$ is a subleading operator of $\Lkb_{n\le k}$, it can receive contributions from \stu-ing $\beta_k \pa^2 H_{k}$, because of the non-linearity of the \stu decompositon.
So to perform an analysis for any subleading term $k>n$ of  $\Lkb_n$
 one would need to use the non-linear \stu decomposition which correctly identifies the propagating degrees of freedom {\it in the decoupling limit} (and not beyond the decoupling limit), as explained in section~\ref{sec:decoupling-limit}. This is the reason why \ref{Prop lin Stuc} is only meaningful at {\it leading order} in a new interaction and beyond its leading order it has to be replaced by \ref{Prop ADM} (which includes \ref{Prop DL}).

 In more physical terms, all the $h_{\mu\nu}$ that appear in the leading order operator $\Lk_n$ may be thought of as living on flat space. Thus it is a real physical degree of freedom living on Minkowski, and so its equations of motion must be second order. Thus there is no requirement that $n_h \leq 1$, for the leading order operator. When dealing with terms at subleading orders, the non-linearities in $h$ should be thought as arising from the curved metric corrections of the leading order term, and so the fields are no longer living on flat space, but rather on the dynamical metric $\eta+h$ and the \stu decomposition does no longer properly identifies the physical degrees of freedom (apart in the decoupling limit, for instance the same issue occurs when dealing with massive gravity on de Sitter \cite{deRham:2012kf}.)\\

%This may seem to run contrary to the philosophy outlined in the beginning of section \ref{sec:decoupling-limit}, where we emphasized the need to work with the non-linear \stu decomposition, and argued that we could only apply the condition of second order equations of motion to interactions within the decoupling limit, consistent with $n_h \leq 1$. But there is no contradiction because here we are asking a different question. There, we were looking to discover the subleading operators of order $k>n$ in $\mathcal{L}^{\rm der}_{n}$, where at leading order (\ie at order $k=n$)$\mathcal{L}^{\rm der}_{n}$ were already satisfying \ref{Prop lin Stuc}. The analysis of {\it subleading} operators requires a non-linear \stu analysis. For instance if $\beta_{k+1}\pa^2 H^{k+1}$ is a subleading operator of $\mathcal{L}^{\rm der}_{n\le k}$, it can receive contributions from \stu-ing $\beta_n \pa^2 H_{n}$, because of the nonlinearity of the \stu decompositon. However when we focus on the leading order operator, non-linear diffeomorphism symmetry reduces trivially to linearized diffs, and we do not need to worry about these corrections. \\

So to summarize, in order for $\Lkb_n$ to have any chance of propagating no more than five dofs, we must first ensure that $\Lk_n$ satisfies the \ref{Prop lin Stuc} and \emph{the equations of motion for $h, A, \pi$ in \eqref{eq:linear-stuck} arising from this operator must be second order in time derivatives}.
%In other words we must have that
%\be
%\label{eq:necessary-condition}
%\frac{\delta}{\delta \Phi^I}\Big[ \beta_n \pa^2 \left(H(\Phi_J)\right)^n \Big] = \mathcal{E}_I(\Phi_J, \pa \Phi_J, \pa^2 \Phi_J),
%\ee
%where $\Phi_I=\{\hat{h}_{\mu\nu}, \hat{A}_\mu, \hat{\pi}\}$.
%
%This may seem to run contrary to the philosophy outlined in the beginning of section \ref{sec:decoupling-limit}, where we emphasized the need to work with the non-linear \stu decomposition, and argued that we could only apply the condition of second order equations of motion to interactions at or below the scale $\Lambda_3$, consistent with $n_h \leq 1$. But there is no contradiction because here we are asking a different question. There, we were looking to discover the subleading operators in $\mathcal{L}^{\rm der}$, and this required a non-linear \stu analysis. The subleading operator $\beta_{n+1}\pa^2 H^{n+1}$ can receive contributions from \stu-ing $\beta_n \pa^2 H_{n}$, because of the nonlinearity of the \stu decompositon. However when we focus on the leading order operator, non-linear diffeomorphism symmetry reduces trivially to linearized diffs, and we don't need to worry about these corrections.

However as we have seen, and as was argued in \cite{Hinterbichler:2013eza}, in four dimensions, up to total derivatives only $\Lk_2$ and $\Lk_3$ satisfy \ref{Prop lin Stuc} and there are no other term $\Lk_n$ with $n\ge4$. As a result there can be no new kinetic interactions that arise at higher order. Furthermore we have seen that $\Lk_3$ has no non-linear completion and is thus not an acceptable term in a theory of gravity. Finally the only acceptable non-linear completion of $\Lk_2$ is the fully non-linear Einstein-Hilbert term. As a result in four dimensions the only possible Lorentz-invariant term with two derivatives which is consistent for a graviton is the Einstein-Hilbert term up to total derivatives.

\subsection{No go for higher order derivative interactions}

Finally we may also use this argument to establish that there are no new interactions at higher orders in derivatives in $D=4$, up to total derivatives (such as the standard Gauss-Bonnet term)\footnote{The higher order Lovelock terms are not present because they vanish identically, not only up to a total derivative, in $D=4$.}. In \cite{Hinterbichler:2013eza} it was argued that the only term in $D=4$ with more than two derivatives that satisfies \ref{Prop lin Stuc} is the linearization of the Gauss-Bonnet term.

We can summarize the argument in \cite{Hinterbichler:2013eza} as follows: the general form of a term satisfying \ref{Prop lin Stuc} with $d$ derivatives and $n_H$ powers of $H$ in $D$ dimensions is, up to total derivatives,
\ba
\mathcal{L}^{\text{(der)}}_{d,n} &=& \ep^{\mu_1 \cdots \mu_{d/2} \nu_1 \cdots \nu'_{D - d/2}} \ep^{\mu'_1 \cdots \mu'_{d/2} \nu'_1 \cdots \nu'_{D-d/2}} \prod_{j=1}^{d/2} \pa_{\mu_j} \pa_{\mu'_j} H_{\nu_j \nu'_j} \prod_{k=d/2 + 1}^{n_H} H_{\nu_k \nu'_k} \prod_{\ell=n_H+1}^{D-d/2} \eta_{\nu_\ell \nu' _\ell}\nn \\
%&=& \ep^{\mu_1 \cdots \mu_{d/2} \nu_1 \cdots \nu_{n_H} \sigma_1 \cdots \sigma_{D-n_H - d/2} } \ep^{\mu'_1 \cdots \mu'_{d/2} \nu'_1 \cdots \nu'_{n_H} \sigma'_1 \cdots \sigma'_{D-n_H-d/2}} \nn \\
%&& \times \prod_{j=1}^{d/2} \pa_{\mu_j} H_{\nu_{2j-1} \nu'_{2j-1}} \pa_{\mu'_j} H_{\nu_{2j} \nu'_{2j}} \prod_{k=d+1}^{n_H} H_{\nu_k \nu'_k} \prod_{\ell = 1}^{D- n_H - d/2} \eta_{\sigma_\ell \sigma'_\ell}  \nn \\
&\sim & \ep \ep (\pa^2 H)^{d/2} H^{n_H-d/2} \eta^{D - n_H - d/2}
\ea
Clearly $d$ must be even for $\mathcal{L}^{\text{(der)}}_{d,n}$ to be a scalar. In order that every derivative acts on an $H$, we must have $n_H \geq d$ (with equality signifying a total derivative), and in order that every index on a derivative or an $H$ be contracted with one of the epsilon tensors we must have $n_H + d/2 \leq D$. For $D=4$ the only possible solution to these constraints with $d>2$ is the total derivative combination with $d=n_H=4$, which is exactly the leading order operator of the Gauss-Bonnet term
\be
\mathcal{L}_{GB} = \ep^{\mu\nu\rho\sigma}\ep_{\mu'\nu'\rho'\sigma'} \pa_\mu \pa^{\mu'} H_\nu^{\nu'} \pa_\sigma \pa^{\sigma'} H_\rho^{\rho'} + O(H^3).
\ee

This rules out any term with $d\geq 4$. For $d=4$, the above term is not the first non-trivial term of a fully non-linear completion, since it is a total derivative. Thus we would need a term with $d=4$ and $n_H>4$ to be the first non-trivial term, and this term would need to satisfy \ref{Prop lin Stuc} (and we know that no such non-total derivative terms exist). Similarly for $d>4$ there are no (non-total derivative) terms that can satisfy \ref{Prop lin Stuc}.

\section{Discussion}
\label{sec:discussion}

The arguments presented in  this paper constitute a no-go theorem stating that new diffeomorphism-violating kinetic terms do not exist in metric formulation of massive gravity, (when maintaining Lorentz invariance). In four dimensions, the only possible term that can involve derivatives for a graviton (be it massive or massless) is the Einstein-Hilbert term. In the context of a massless-spin-2 field this statement was already well-known and was deduced using diffeomorphism invariance \cite{Gupta:1954zz,Weinberg:1965rz,Deser:1969wk,Feynman:1996kb,Boulware:1974sr}.
We have now proven the equivalent for massive gravity. The unique theory for a massive spin-2 field in four dimensions is
\ba
S_{\rm mGR}= \int\d^4 x\(\sqrt{-g}\frac{\mpl^2}{2}R-\sqrt{-g}\Lambda +\frac{\mpl^2m^2}{4}\sum_{n=2}^4 \alpha_n \nn \Lmb_{n}+\sqrt{-g}\L_{\rm matter}(g, \psi)\)\,,
\ea
where we have included a cosmological constant $\Lambda$ and coupling to matter $\L_{\rm matter}$. \\

This most general action for a graviton was derived assuming that there are no more than five propagating degrees of freedom in the graviton. Interestingly, we find that the only possible derivative term is nothing other than the standard Einstein-Hilbert term which by itself only propagates two degrees of freedom (it would be the mass term which would be responsible for the propagation of the three additional degrees of freedom in massive gravity).
Thus we have recovered the uniqueness of the Einstein-Hilbert term by requiring a much weaker condition than what is assumed in General Relativity. This could have important consequences for the quantum stability of a theory of massive gravity (see Refs.~\cite{deRham:2012ew,deRham:2013qqa}).\\

This result can be paralleled with that of Refs.~\cite{Khoury:2011ay,Khoury:2013oqa} where it was proven that even if one breaks Lorentz invariance the only theory which remains spatially covariant and does not propagate more than two degrees of freedom is General Relativity.  These considerations add up to the realization that General Relativity and particularly the Einstein-Hilbert term is extremely unique and special even beyond the requirement of diffeomorphism invariance that is used in General Relativity. In other words we do not need to assume diffeomorphism invariance to be led to the central importance of the Riemann curvature in a theory of gravity.\\

The sets of possibilities considered in this paper were first motivated by Gauss-Bonnet gravity and the fact that it is ghost-free \cite{Lovelock:1971yv}. However upon failure of this approach we have considered the {\it most general} sets of derivative interactions which
\begin{enumerate}%[label=(\alpha{*}), ref=(\alpha{*})]
  \setlength{\itemsep}{0pt}
   \setlength{\parskip}{0pt}
   \setlength{\parsep}{0pt}
\item[(a)] Respect Lorentz invariance.
\item[(b)] Are local and connect to flat space-time (\ie admit flat space vacuum solution in the absence of a cosmological constant and matter) with no linearization instability\footnote{The latter rules out the often quoted exception of $f(R)$ models. These models, which are better interpreted as gravity and a scalar field, admit a linearization instability around Minkowski space-time since perturbatively there would appear to be two degrees of freedom, but non-perturbatively there are three.}.
\item[(c)] Can be formulated with a metric.
\item[(d)] Do not propagate more than five degrees of freedom.
\end{enumerate}
Most of these requirements are intrinsic to the very definition of the notion of a spin-2 field and it is therefore difficult to avoid them. It is also hard to imagine terms with less structure containing the necessary constraints to remove unwanted degrees of freedom. \\

We should emphasize that the difficulty we have found is the standard tension that arises from demanding both Lorentz invariance and a theory propagating five degrees of freedom. We could easily find sensible Lorentz-violating kinetic terms for massive gravity (although not for massless gravity \cite{Khoury:2011ay,Khoury:2013oqa}). \\

Another possible way out of these results would be to add additional ghost-free degrees of freedom to the action.  One such possibility could be accounted for by allowing for the existence of torsion. We worked directly with the metric, however really we should work directly with the vielbein when discretizing. The deconstruction approach we followed amounts to assuming that the five-dimensional torsion-free expression for the spin connection was satisfied before discretizing. It would be interesting to discretize in a first order form where $\Om^{AB}$ was kept as an independent variable, and solve the equations after discretizing. If one is able to integrate out the connection, we should be able to get back to the metric formulation and the results provided in the paper would hold. If the degrees of freedom in the connection cannot be integrated out without leading to non-localities, then the theory would genuinely have more than five degrees of freedom. However it is possible that these degrees of freedom could be healthy. This is a direction that could be explored.  \\

Finally we have been working in the metric language which is not always equivalent to the vielbein formulation of massive gravity~\cite{Deffayet:2012zc}. However when working with Minkowski as a reference metric and assuming trivial vacuum solutions, the perturbative analysis we have performed can be carried out in both formulations without any differences.

\acknowledgments

We would like to thank Lavinia Heisenberg and Nick Ondo for useful discussions. AJT is supported by Department of Energy Early Career Award DE-SC0010600. AM is supported by an NSF Graduate Research Fellowship. CdR is supported by a Department of Energy grant DE-SC0009946.

\appendix
\section{Details of decoupling limit analysis}
\subsection{Cubic order}
In this subsection we will give a more explicit explanation of the analysis at cubic order. We start with the most general Lagrangian of the form $\pa^2 H^3$, which after integrating by parts and removing total derivative combinations can be put in the form
\ba
\mathcal{L}^{(3)}_{\rm gen} &=& b_1 H^{\mu\nu}\pa_\mu H^{\rho\sigma}\pa_\nu H_{\rho\sigma} + b_2 H^{\mu\nu} \pa_\mu H^\sigma_{\ \ \nu}\pa_\sigma H + b_3 H^{\mu\nu} \pa_\mu H \pa_\nu H  \nn \\
&+& b_4 H^{\mu\nu} \pa_\sigma H \pa^\sigma H_{\mu\nu} + b_5 H \pa_\mu H \pa^\mu H + b_6 H^{\mu\nu} \pa_\rho H_{\mu \sigma}\pa^\sigma H^\rho_{\ \ \nu} \nn \\
&+& b_7 H \pa_\mu H_{\nu\sigma} \pa^\sigma H^{\mu\nu} + b_8 H^{\mu\nu} \pa_\rho H_{\mu \sigma}\pa^\rho H^\sigma_{\ \ \nu} + b_9 H \pa_\mu H_{\nu \sigma} \pa^{\mu} H^{\nu\sigma} \nn \\
&+& b_{10} H^{\mu\nu} \pa_\sigma H^\sigma_{\ \ \mu} \pa_\rho H^\rho_{\ \ \nu} + b_{11} H^{\mu\nu} \pa_\mu H^\rho_{\ \ \nu} \pa_\sigma H^\sigma_{\ \ \rho} + b_{12} H \pa_\mu H^{\mu\nu}\pa_\sigma H^\sigma_{\ \ \nu} \nn \\
&+& b_{13} H^{\mu\nu} \pa^\rho H_{\mu\nu}\pa^\sigma H_{\rho \sigma}+ b_{14} H \pa^\mu H \pa^\nu H_{\mu\nu}.
\ea
All indices are raised and lowered with $\eta_{\mu\nu}$. \\

We then perform the nonlinear \stu decomposition, given by \eqref{eq:non-linear-stuck}, on the whole Lagrangian  $\L=a \mathcal{L}_a^{(2)} + \mathcal{L}^{(3)}_{\rm gen}$. We only keep interactions up to and including the scale $\Lambda_3$. This amounts to keeping terms that are
\begin{itemize}
\item Zeroth order in $A$ and $h$
\item First or Second order in $A$ and Zeroth order in $h$
\item First order in $h$ and Zeroth order in $A$.
\end{itemize}
We vary the lagrangian and demand that all equations contain at most two derivatives. This yields a set of constraints on the parameters. As an example, when varying the Lagrangian with respect to $\pi$ we find the term
\be
\frac{\delta \mathcal{L}}{\delta \pi}\supset - \frac{4}{\Lambda_3^5} \Big[a + 4 b_3 + 2 b_7 + 2 b_9\Big] \pa_\mu \pa_\nu \pa_\sigma h \pa^\mu \pa^\nu \pa^\sigma \pi .
\ee
Demanding that this term vanishes puts a constraint on the parameters $b_j$. Note that the non-linearity of the \stu decomposition forces the $b_j$ coefficients to depend on $a$. \\

The system of constraints was found and solved using the \emph{Mathematica} package xAct \cite{xAct}. After eliminating all interactions with higher derivative equations of motion, the result is a three parameter family of cubic lagrangians that are ghost free up to the scale $\Lambda_3$ at cubic order.
\ba
\label{eq:L3Lambda3}
\mathcal{L}^{(3)}_{{\rm g.f.},\Lambda_3} &=& a \left( \mathcal{L}^{(2)}_a + \mathcal{L}^{(3)}_a \right)  + b_1 \mathcal{L}^{(3)}_{b_1} + b_2 \mathcal{L}^{(3)}_{b_2} \nn \\
&=& -2a H^{\mu\nu}  \Ein^{\alpha\beta}\mn H_{\alpha\beta} \nn \\
&+& b_1 H^{\mu\nu}\pa_\mu H^{\rho\sigma}\pa_\nu H_{\rho\sigma}
+ b_2 H^{\mu\nu} \pa_\mu H^\sigma_{\ \ \nu}\pa_\sigma H
- b_1 H^{\mu\nu} \pa_\mu H \pa_\nu H \nn \\
&+& \left( 2b_1 - b_2 \right) H^{\mu\nu} \pa_\sigma H \pa^\sigma H_{\mu\nu}
+ \left( -b_1 +\frac{1}{2}b_2 + \frac{a}{2} \right)H \pa_\mu H \pa^\mu H \nn \\
&+&  \left( 5b_1 - b_2 -a \right) H^{\mu\nu} \pa_\rho H_{\mu \sigma}\pa^\sigma H^\rho_{\ \ \nu}
+  \left( b_1 +\frac{1}{2} b_2 \right) H \pa_\mu H_{\nu\sigma} \pa^\sigma H^{\mu\nu} \nn \\
&+&  \left( -2b_1 + b_2 \right) H^{\mu\nu} \pa_\rho H_{\mu \sigma}\pa^\rho H^\sigma_{\ \ \nu}
+  \left( b_1 - \frac{1}{2}b_2 - \frac{1}{2} a \right) H \pa_\mu H_{\nu \sigma} \pa^{\mu} H^{\nu\sigma} \nn \\
&+&  \left( -3b_1 + a \right) H^{\mu\nu} \pa_\sigma H^\sigma_{\ \ \mu} \pa_\rho H^\rho_{\ \ \nu}
 - b_2 H^{\mu\nu} \pa_\mu H^\rho_{\ \ \nu} \pa_\sigma H^\sigma_{\ \ \rho}  \nn \\
&+&  \left( -3 b_1 + \frac{1}{2} b_2 + a\right) H \pa_\mu H^{\mu\nu}\pa_\sigma H^\sigma_{\ \ \nu}
+  \left( -2b_1 + b_2 \right) H^{\mu\nu} \pa^\rho H_{\mu\nu}\pa^\sigma H_{\rho \sigma}  \nn \\
&+&  \left( 2b_1 - b_2 - a \right) H \pa^\mu H \pa^\nu H_{\mu\nu} .
\ea
This equation also serves to define $\mathcal{L}^{(3)}_a, \mathcal{L}^{(3)}_{b_1}, \mathcal{L}^{(3)}_{b_3}$ which appear in the main text,
\ba
\label{L3a}
\mathcal{L}^{(2)}_a + \mathcal{L}^{(3)}_a&=&\mathcal{L}^{(3)}_{{\rm g.f.},\Lambda_3}\Big|_{a=1, b_1=0, b_2=0}\\
\label{L3b1}
\L^{(3)}_{b_1}&=&\mathcal{L}^{(3)}_{{\rm g.f.},\Lambda_3}\Big|_{a=0, b_1=1, b_2=0}\\
\L^{(3)}_{b_2}&=&\mathcal{L}^{(3)}_{{\rm g.f.},\Lambda_3}\Big|_{a=0, b_1=0, b_2=1}\,.
\label{L3b2}
\ea
This \stu analysis is not a proof that the terms are ghost free. Indeed an ADM analysis described in the main text shows that only the combination of parameters corresponding to the Einstein-Hilbert term is ghost free. Rather, this is a necessary condition.

\subsection{Quartic order}
We follow the same analysis described in the previous section. For completeness and to fix the notation we show the important results. The most general Lagrangian of the form $\pa^2 H^4$, after removing total derivatives, can be written in the form
\ba
\label{eq:L4gen}
\mathcal{L}^{(4)}_{\rm gen} &=&
c_1 H^{\alpha \beta}H^{\gamma}_{\ \ \alpha} \pa_\beta H^{\delta \epsilon} \pa_\gamma H_{\delta \epsilon}
+ c_2 H^\gamma_{\ \ \alpha} H^{\alpha \beta} \pa_\gamma H^\delta_{\ \ \beta} \pa_\delta H
+ c_3 H^\gamma_{\ \ \alpha} H^{\alpha \beta} \pa_\beta H \pa_\gamma H \nn \\
&+& c_4 H H^{\beta \gamma} \pa_\beta H \pa_\gamma H
+ c_5 H^{\alpha \beta} H^{\gamma \delta} \pa_\gamma H^\epsilon_{\ \ \alpha} \pa_\delta H_{\beta \epsilon}
+ c_6 H^{\alpha \beta} H^{\gamma \delta} \pa_\beta H^\epsilon_{\ \ \alpha} \pa_\delta H_{\gamma \epsilon} \nn \\
&+& c_7 H^{\alpha \beta} H^{\gamma \delta} \pa_\beta H_{\alpha \gamma} \pa_\delta H
+ c_8 H^{\alpha \beta} H^{\gamma \delta} \pa_\gamma H_{\alpha \beta} \pa_\delta H
+ c_9 H H^{\beta \gamma}\pa_{\beta}H^{\delta \epsilon} \pa_{\gamma}H_{\delta \epsilon}  \nn \\
&+& c_{10} H H^{\beta \gamma} \pa_\gamma H^\delta_{\ \ \beta} \pa_\delta H
+ c_{11} H^\gamma_{\ \ \alpha} H^{\alpha \beta} \pa_\delta H \pa^\delta H_{\beta \gamma}
+ c_{12} H H^{\beta \gamma} \pa_\delta H \pa^\delta H_{\beta \gamma}\nn \\
&+&c_{13} H_{\alpha \beta}H^{\alpha \beta} \pa_\delta H \pa^\delta H
+ c_{14} H^2 \pa_\alpha H \pa^\alpha H
+ c_{15} H H^{\beta \gamma} \pa_\delta H^\delta_{\ \ \beta} \pa_\epsilon H^\epsilon_{\ \ \gamma} \nn \\
&+& c_{16} H^{\alpha \beta} H^{\gamma \delta} \pa_\beta H_{\alpha \gamma}\pa_\epsilon H^\epsilon_{\ \ \delta}
+ c_{17} H^{\gamma}_{\ \ \alpha} H^{\alpha \beta} \pa_\gamma H^\delta_{\ \ \beta}\pa_\epsilon H^\epsilon_{\ \ \delta}
+ c_{18} H H^{\beta \gamma} \pa_\gamma H^\delta_{\ \ \beta}\pa_\epsilon H^\epsilon_{\ \ \delta} \nn \\
&+& c_{19} H_{\alpha \beta}H^{\alpha \beta} \pa_\gamma H^{\gamma \delta} \pa_\epsilon H^\epsilon_{\ \ \delta}
+ c_{20} H^2 \pa_\gamma H^{\gamma \delta}\pa_\epsilon H^\epsilon_{\ \ \delta}
+ c_{21} H^\gamma_{\ \ \alpha} H^{\alpha \beta} \pa^\delta H_{\beta \gamma} \pa_\epsilon H^\epsilon_{\ \ \delta} \nn \\
&+& c_{22} H H^{\beta \gamma} \pa^\delta H_{\beta \gamma} \pa_\epsilon H^\epsilon_{\ \ \delta}
+ c_{23} H_{\alpha \beta}H^{\alpha \beta} \pa^\delta H \pa_\epsilon H^\epsilon_{\ \ \delta}
+ c_{24} H^2 \pa^\delta H \pa_\epsilon H^\epsilon_{\ \ \delta} \nn \\
&+& c_{25} H^{\alpha \beta}H^{\gamma \delta} \pa_\delta H_{\gamma \epsilon} \pa^\epsilon H_{\alpha \beta}
+ c_{26} H^{\alpha \beta} H^{\gamma \delta} \pa_{\epsilon} H_{\gamma \delta}\pa^\epsilon H_{\alpha \beta}
+ c_{27} H^{\alpha \beta} H^{\gamma \delta} \pa_\delta H_{\beta \epsilon}\pa^\epsilon H_{\alpha \gamma} \nn \\
&+& c_{28} H^{\alpha \beta} H^{\gamma \delta} \pa_\epsilon H_{\beta \delta} \pa^\epsilon H_{\alpha \gamma}
+ c_{29} H^\gamma_{\ \ \alpha} H^{\alpha \beta} \pa_\gamma H_{\delta \epsilon} \pa^\epsilon H^\delta_{\ \ \beta}
+ c_{30} H H^{\beta \gamma} \pa_\gamma H_{\delta \epsilon} \pa^\epsilon H^\delta_{\ \ \beta} \nn \\
&+& c_{31} H^\gamma_{\ \ \alpha} H^{\alpha \beta} \pa_{\delta} H_{\gamma \epsilon} \pa^\epsilon H^\delta_{\ \ \beta}
+ c_{32} H H^{\beta \gamma} \pa_\delta H_{\gamma \epsilon} \pa^\epsilon H^\delta_{\ \ \beta}
+ c_{33} H^\gamma_{\ \ \alpha} H^{\alpha \beta} \pa_\epsilon H_{\gamma \delta} \pa^\epsilon H^\delta_{\ \ \beta} \nn \\
&+& c_{34} H H^{\beta \gamma} \pa_\epsilon H_{\gamma \delta} \pa^\epsilon H^\delta_{\ \ \beta}
+ c_{35} H_{\alpha \beta} H^{\alpha \beta} \pa_\delta H_{\gamma \epsilon} \pa^{\epsilon} H^{\gamma \delta}
+ c_{36} H^2 \pa_{\delta} H_{\gamma \epsilon} \pa^\epsilon H^{\gamma \delta} \nn \\
&+& c_{37} H_{\alpha \beta} H^{\alpha \beta} \pa_\epsilon H_{\gamma \delta} \pa^\epsilon H^{\gamma \delta}
+ c_{38} H^2 \pa_\epsilon H_{\gamma \delta} \pa^\epsilon H^{\gamma \delta}.
\label{L4Gen}
\ea

By applying the nonlinear \stu analysis as described in the previous section, we obtain 36 constraints on the 38 parameters, that can be solved as expressed  in Table~\ref{relation between the c's}.

\renewcommand\thetable{sol.c}
\begin{table}[h]
\caption{Relations between the $c$ coefficients so that $\mathcal{L}^{(4)}$ in \eqref{L4Gen} satisfies \ref{Prop DL}.}
\label{relation between the c's}
\centering
\renewcommand{\arraystretch}{1.3}
\begin{tabular}{c | c}
%% Column 1
$\begin{array} {lcl}
c_3 &=&- c_1 \nn \\
c_4 &=& \frac{1}{4}\left(-a + 9 b_1 - 2 b_2 + 4 c_1\right) \nn \\
c_5 &=&  \frac{1}{2}\left( -a + 7b_1 - 2b_2 +4 c_1 \right)  \nn \\
c_6 &=& \frac{1}{2}\left(a - 7 b_1 + 2b_2 - 4 c_1\right) \nn \\
c_7 &=& \frac{1}{4}\left( -2a + 18 b_1 - 5b_2 + 8c_1 \right) \nn \\
c_8 &=& \frac{1}{2}\left(a - 7b_1 + 2b_2 - 4c_1\right) \nn \\
c_9 &=& \frac{1}{4}\left(a - 9b_1 + 2 b_2 - 4c_1\right)  \nn \\
c_{10}&=&a - 5b_1 + b_2 - c_2  \nn \\
c_{11}&=&\frac{1}{4}\left(- 4 b_1 + b_2 + 8 c_1 - 4 c_2\right)  \nn \\
c_{12}&=&\frac{1}{2}\left(-a + b_1 - 4c_1 + 2c_2\right)  \nn \\
c_{13}&=&\frac{1}{8} \left(-a + b_1 - 4 c_1 + 2 c_2 \right)  \nn \\
c_{14}&=& \frac{1}{8} \left(b_1 - b_2 + 4 c_1 - 2 c_2\right)\nn \\
c_{15}&=& \frac{1}{4}\left(-3a + 13 b_1 - 3b_2 - 4 c_1 \right) \nn \\
c_{16}&=&  \frac{1}{4}\left(5a - 35 b_1 + 10 b_2 - 16 c_1 + 2 c_2 \right) \nn \\
c_{17}&=& \frac{1}{4}\left( a + 9 b_1 - 4b_2 + 24 c_1 - 6 c_2 \right)\nn \\
c_{18}&=&  \frac{1}{4}\left( -2a + 2 b_1 + b_2 - 16 c_1 + 4 c_2 \right)\nn \\
c_{19}&=&  \frac{1}{8} \left( a - 15 b_1 + 5 b_2 - 12 c_1 + 2 c_2 \right) \nn \\
c_{20}&=& \frac{1}{8}\left( -3a + 21 b_1 - 6 b_2 + 12 c_1 - 2 c_2\right) \nn \\
\end{array}$
& %% Column 2
$\begin{array} {lcl}
c_{21}&=& b_1 - \frac{1}{4} b_2 - 2 c_1 + c_2 \nn \\
c_{22}&=& \frac{1}{2} \left( a - b_1 + 4 c_1 - 2 c_2 \right) \nn \\
c_{23}&=& \frac{1}{4}\left( a - b_1 + 4 c_1 - 2 c_2 \right) \nn \\
c_{24}&=&  \frac{1}{4}\left( -b_1 + b_2 - 4 c_1 + 2 c_2 \right)\nn \\
c_{25}&=& -a + 5 b_1 - \frac{3}{2} b_2 + c_2 \nn \\
c_{26}&=&\frac{1}{4}\left( a - 3 b_1 + b_2 + 4c_1 - 2 c_2 \right) \nn \\
c_{27}&=& \frac{1}{4}\left( -a + 7 b_1 - 2 b_2 + 16 c_1 - 6 c_2 \right)\nn \\
c_{28}&=&\frac{1}{4}\left( -a + 7 b_1 - 2b_2 - 4 c_1 + 2 c_2 \right) \nn \\
c_{29}&=&-\frac{1}{4} a - \frac{9}{4} b_1 + b_2 - 6 c_1 + \frac{1}{2} c_2 \nn \\
c_{30}&=&\frac{1}{4} \left( -2a + 18 b_1 - 5 b_2 + 16 c_1 \right) \nn \\
c_{31}&=& 2c_1 - c_2\nn \\
c_{32}&=&\frac{1}{4}\left(a - 11b_1 + 3 b_2 - 4 c_1 + 4 c_2\right) \nn \\
c_{33}&=& -2 c_1 + c_2\nn \\
c_{34}&=&  \frac{1}{2} \left(a - b_1 + 4 c_1 - 2c_2 \right)\nn \\
c_{35}&=& \frac{1}{8}\left(-3a + 17 b_1 - 5b_2 + 4 c_1 + 2 c_2\right) \nn \\
c_{36}&=& \frac{1}{8}\left( 3a - 19 b_1 + 4 b_2 - 4c_1 - 2 c_2 \right) \nn \\
c_{37}&=&\frac{1}{8}\left( a- b_1 + 4 c_1 - 2 c_2 \right) \nn \\
c_{38}&=& \frac{1}{8}\left(-b_1 + b_2 - 4c_1 + 2c_2\right)\nn \\
\end{array}$ \\
\end{tabular}
\end{table}
We define the two `new' quartic Lagrangians which satisfy \ref{Prop DL} as the ones proportional to $c_1$ and $c_2$ respectively,
\ba
\label{L4c1}
\L^{(4)}_{c_1}&=&\mathcal{L}^{(4)}_{\rm gen}\Big|_{{\rm \ref{relation between the c's}},\ a=0, b_1=0, b_2=0, c_1=1, c_2=0}\\
\L^{(4)}_{c_2}&=&\mathcal{L}^{(4)}_{\rm gen}\Big|_{{\rm \ref{relation between the c's}},\ a=0, b_1=0, b_2=0, c_1=0, c_2=1}\,,
\label{L4c2}
\ea
and the extension of the previously found quadratic and cubic Lagrangians,
\ba
\label{L4a}
\L^{(4)}_{a}&=&\mathcal{L}^{(4)}_{\rm gen}\Big|_{{\rm \ref{relation between the c's}},\ a=1, b_1=0, b_2=0, c_1=0, c_2=0}\\
\L^{(4)}_{b_1}&=&\mathcal{L}^{(4)}_{\rm gen}\Big|_{{\rm \ref{relation between the c's}},\ a=0, b_1=1, b_2=0, c_1=0, c_2=1}\,,
\label{L4b1}\\
\L^{(4)}_{b_2}&=&\mathcal{L}^{(4)}_{\rm gen}\Big|_{{\rm \ref{relation between the c's}},\ a=0, b_1=0, b_2=1, c_1=0, c_2=1}\,,
\label{L4b2}
\ea
We also define the full Lagrangians up to quartic order
\ba
\label{La}
\L_a&=&\mathcal{L}^{(2)}_a + \mathcal{L}^{(3)}_a+\mathcal{L}^{(4)}_a\\
\label{Lb1}
\L_{b_1}&=&\mathcal{L}^{(3)}_{b_1}+\mathcal{L}^{(4)}_{b_1}\\
\label{Lb2}
\L_{b_2}&=&\mathcal{L}^{(3)}_{b_2}+\mathcal{L}^{(4)}_{b_2}\\
\label{Lc1}
\L_{c_1}&=&\L^{(4)}_{c_1}\,.
\ea

\pagebreak
%%%%%%%%%%%%%%%%%%%%%%%%%%%%%%%%%%%%%%%%%%%%%%%%%%%%%%%%%%%%%%%%%%%%%
%%%% Bibliography

%\newpage
\bibliographystyle{JHEPmodplain}
\bibliography{refs_2}

\end{document}